\documentclass[aps,prx,reprint,showpacs,twocolum]{revtex4-2}
\usepackage{amsmath,amssymb,graphicx}
\usepackage{amsmath,amssymb,graphicx}
\usepackage{multirow}
\usepackage{booktabs}
\usepackage{makecell}
\usepackage{color}
\usepackage{float}
\usepackage{titletoc}
\newcommand{\bitem}{\begin{itemize}}
\newcommand{\fitem}{\end{itemize}}
\newcommand{\beq}{\begin{equation}}
\newcommand{\eeq}{\end{equation}}
\newcommand{\beqa}{\begin{eqnarray}}
\newcommand{\eeqa}{\end{eqnarray}}

\newcommand{\ket}[1]{\ensuremath{|#1\rangle}}

\begin{document}

	\title{Experimental characterization of quantum many-body
		localization transition}
	
	\author{Ming Gong$^{1,2,3}$}
	\thanks{These authors contributed equally to this work.}
	\author{Gentil D. de Moraes Neto$^{4}$}
	\thanks{These authors contributed equally to this work.}
	\author{Chen Zha$^{1,2,3}$}
	\author{Yulin Wu$^{1,2,3}$}
	\author{Hao Rong$^{1,2,3}$}
	\author{Yangsen Ye$^{1,2,3}$}
	\author{Shaowei Li$^{1,2,3}$}
	\author{Qingling Zhu$^{1,2,3}$}
	\author{Shiyu Wang$^{1,2,3}$}
	\author{Youwei Zhao$^{1,2,3}$}
	\author{Futian Liang$^{1,2,3}$}
	\author{Jin Lin$^{1,2,3}$}
	\author{Yu Xu$^{1,2,3}$}
	\author{Cheng-Zhi Peng$^{1,2,3}$}
	\author{Hui Deng$^{1,2,3}$}
	\author{Abolfazl Bayat$^{4}$}
	\email{abolfazl.bayat@uestc.edu.cn}
	\author{Xiaobo Zhu$^{1,2,3}$}
	\email{xbzhu16@ustc.edu.cn}
	\author{Jian-Wei Pan$^{1,2,3}$}
	\affiliation{$^1$  Hefei National Laboratory for Physical Sciences at the Microscale and Department of Modern Physics, University of Science and Technology of China, Hefei 230026, China}
	\affiliation{$^2$  Shanghai Branch, CAS Center for Excellence in Quantum Information and Quantum Physics, University of Science and Technology of China, Shanghai 201315, China}
	\affiliation{$^3$  Shanghai Research Center for Quantum Sciences, Shanghai 201315, China}
	\affiliation{$^{4}$Institute of Fundamental and Frontier Sciences, University of Electronic Science and Technology of China, Chengdu 610054, China}
	
\begin{abstract}
As strength of disorder enhances beyond a threshold value in many-body systems, a fundamental transformation happens through which the entire spectrum localizes, a phenomenon known as many-body localization. This has profound implications as it breaks down fundamental principles of statistical mechanics, such as thermalization and ergodicity. Due to the complexity of the problem, the investigation of the many-body localization transition has remained a big challenge. The experimental exploration of the transition point is even more challenging as most of the proposed quantities for studying such effect are practically infeasible. Here, we experimentally implement a scalable protocol for detecting the many-body localization transition point, using the dynamics of a $N=12$ superconducting qubit array. We show that the sensitivity of the dynamics to random samples becomes maximum at the transition point which leaves its fingerprints in all spatial scales. By exploiting three quantities, each with different spatial resolution, we identify the transition point with excellent match between simulation and experiment. In addition, one can detect the evidence of mobility edge through slight variation of the transition point as the initial state varies. The protocol is easily scalable and can be performed across various physical platforms. 
\end{abstract}	

\maketitle	
	
\begin{figure*}
		\centering \includegraphics[width=1\linewidth]{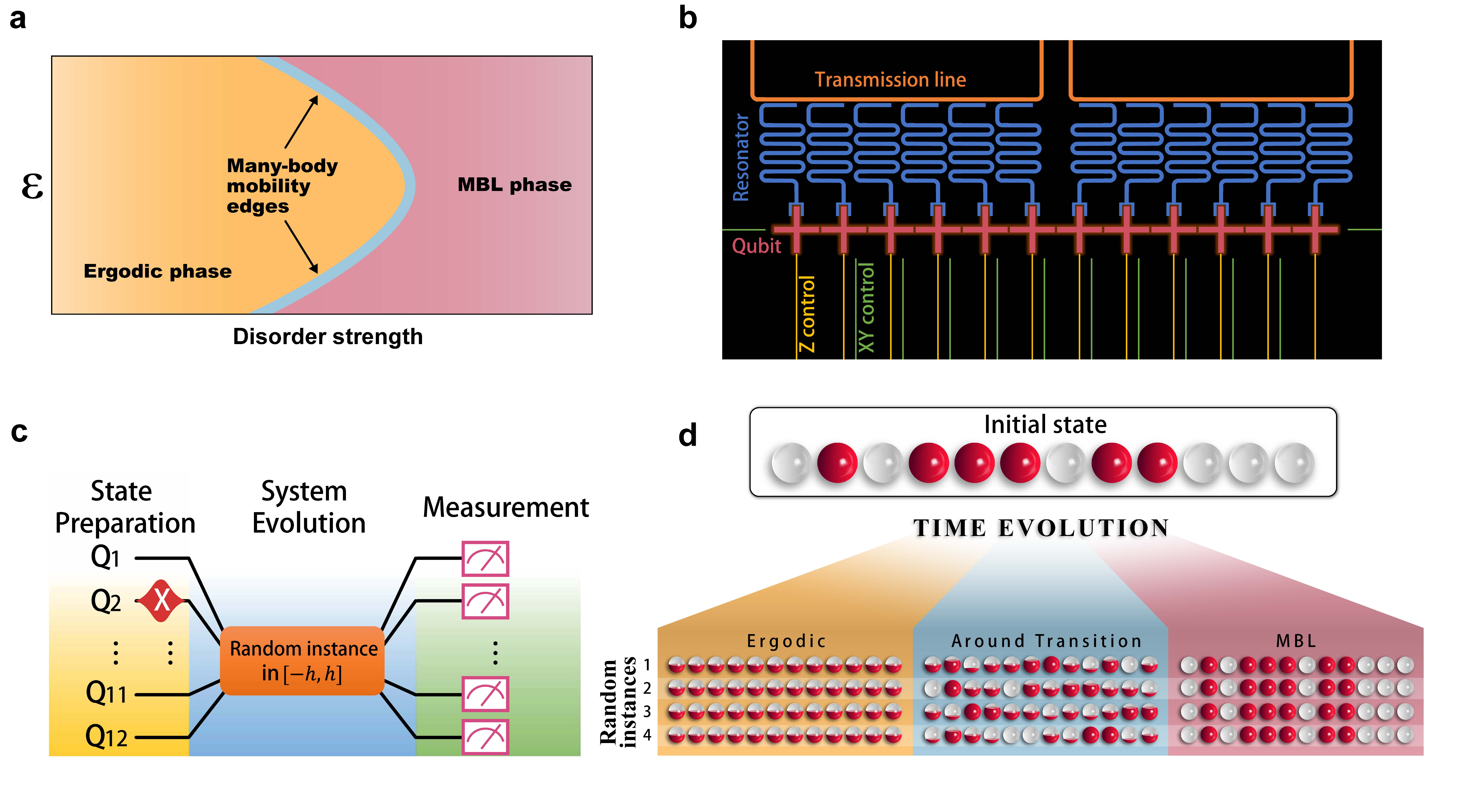}
		\caption{\textbf{Schematics of the experiment.} \textbf{a}, The illustration of the phase diagram of the system across the whole energy spectrum versus disorder strength. Each energy eigenstates localizes at a different disorder strength resulting in many-body mobility edge. \textbf{b}, The twelve transmon qubits, illustrated as red crosses, are arranged in a 1D chain. The direct coupling between them is realized via capacitors. Each qubit has individual $Z$ (yellow) and $XY$ (green) control lines for state manipulation. The twelve readout resonators (blue) are dispersively coupled to their corresponding qubits, and then divided into two groups to couple to the transmission lines (orange). \textbf{c}, The three steps of the experimental procedure are depicted, namely initialization, evolution and readout. For state preparation, the corresponding qubits are excited to $\ket{1}$ by applying $X$ gates. After that, for the state evolution, all the qubits are tunned to their working frequencies $\omega+h_{\ell}$ with $h_{\ell}$ being a random number in the interval $[-h,h]$. After time $t$, for readout, the qubits are detuned to stop the evolution and then simultaneously  measured in $\sigma_z$ direction. \textbf{d}, Our protocol can be understood in a simple way. Each circle represents a qubit and the filling color shows the corresponding average population. Namely, empty, full and fractional filling colors stand for $\left\langle \widehat{n_{\ell }}\right\rangle =0$, $\left\langle \widehat{n_{\ell }}\right\rangle=1$ and $0<\left\langle \widehat{n_{\ell }}\right\rangle<1$, respectively. The dynamics start with a given initial state. Each column shows the average population of each site at long time dynamics for a specific random instance of qubit frequencies. In the ergodic regime, the system thermalizes with $\left\langle \widehat{n_{\ell }}\right\rangle \sim 1/2$ and thus the variation with respect to different random realizations is small. Deep in the MBL regime the dynamics is almost frozen and $\left\langle \widehat{n_{\ell }}\right\rangle$ remains close to its initial value leading also to weak dependence to the different random realizations. Remarkably, around the MBL transition, the long time dynamics shows a strong dependence on the random potential. We exploit this feature to identify the MBL transition point.}
		\label{fig:fig1}
	\end{figure*}
	
	\begin{figure*}
		\centering \includegraphics[width=1.\linewidth]{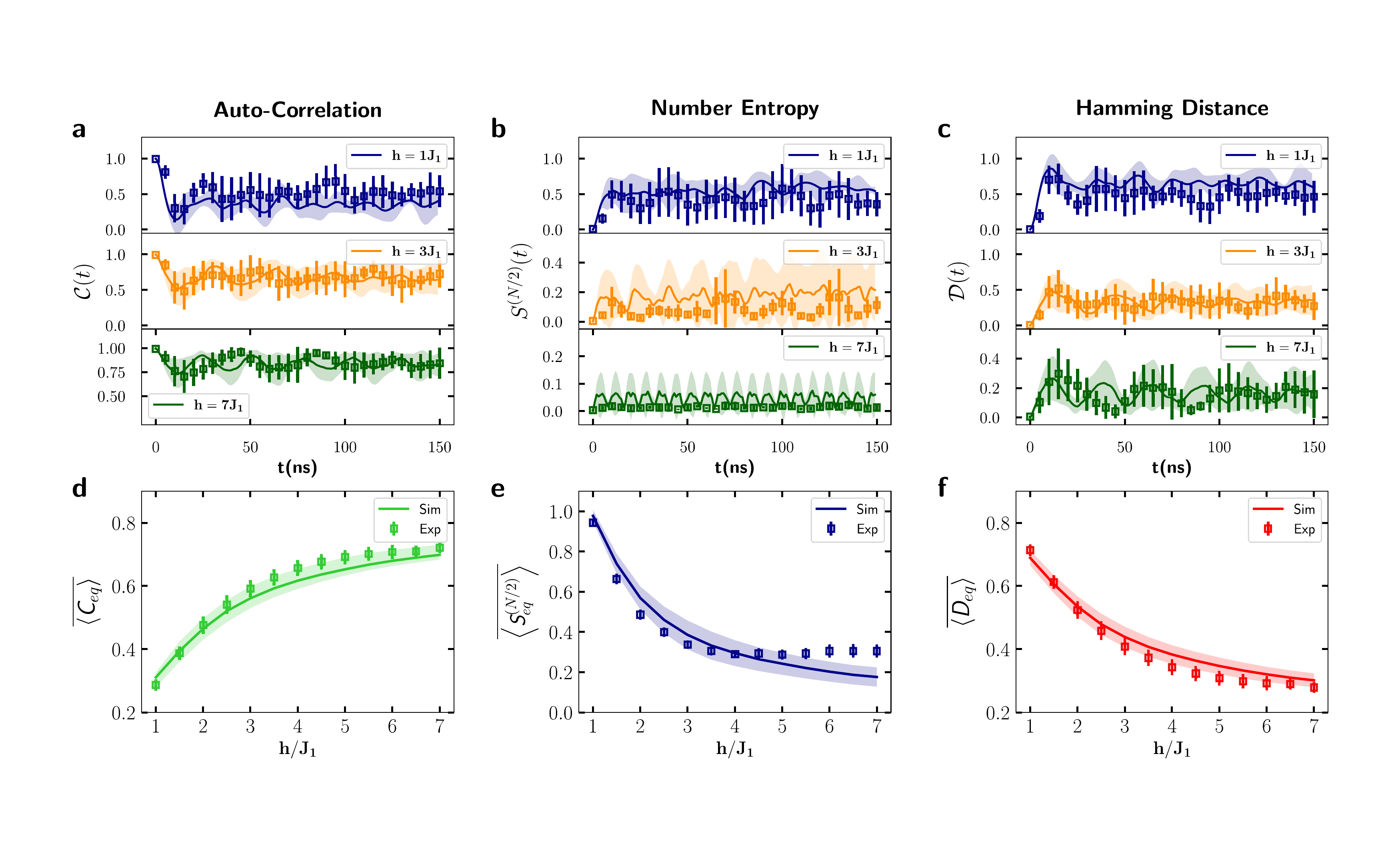}
		\caption{\textbf{Dynamics of the quantities and equilibrium values.} The non-equilibrium dynamics of the system, initialized in a Neel state $\left\vert 0,1,\cdots,1\right\rangle$, averaged over five random instances, for the three quantities: \textbf{a}, the auto-correlation $\mathcal{C}(t)$; \textbf{b}, the number entropy of half-chain $S^{(N/2)}(t)$ ; and \textbf{c},the Hamming distance $\mathcal{D}(t)$. The solid lines are numerical simulations and markers are the experimental data. The disorder strength $h$ is chosen to be $h/J_1=1$ (ergodic), $h/J_1=3$ (near the MBL transition) and  $h/J_1=7$ (MBL).  All the three quantities  equilibrate after a transient time. To see the long time behavior, we consider the equilibrium values averaged over all given initial states and $60$ random realizations as a function of disorder strength $h/J_1$ for: \textbf{d}, the auto-correlation function $\overline{\left\langle \mathcal{C}_{eq}\right\rangle}$; \textbf{e}, the number entropy of half chain $\overline{\left\langle S_{eq}^{(N/2)}\right\rangle }$; and \textbf{f}, the Hamming distance $\overline{\left\langle \mathcal{D}_{eq}\right\rangle }$. Again, the solid lines are numerical simulations and markers are the experimental data. All the three quantities smoothly changes from the ergodic to the MBL phase without revealing the MBL transition point. The standard deviation, depicted as shadow for numerical simulation and error bars for experimental data, quantifies the uncertainty in our estimations. We note a good agreement between simulation and experiment. 
		}
		\label{fig:fig2}
	\end{figure*}	
	
	\begin{figure*}
		\includegraphics[width=1.\linewidth]{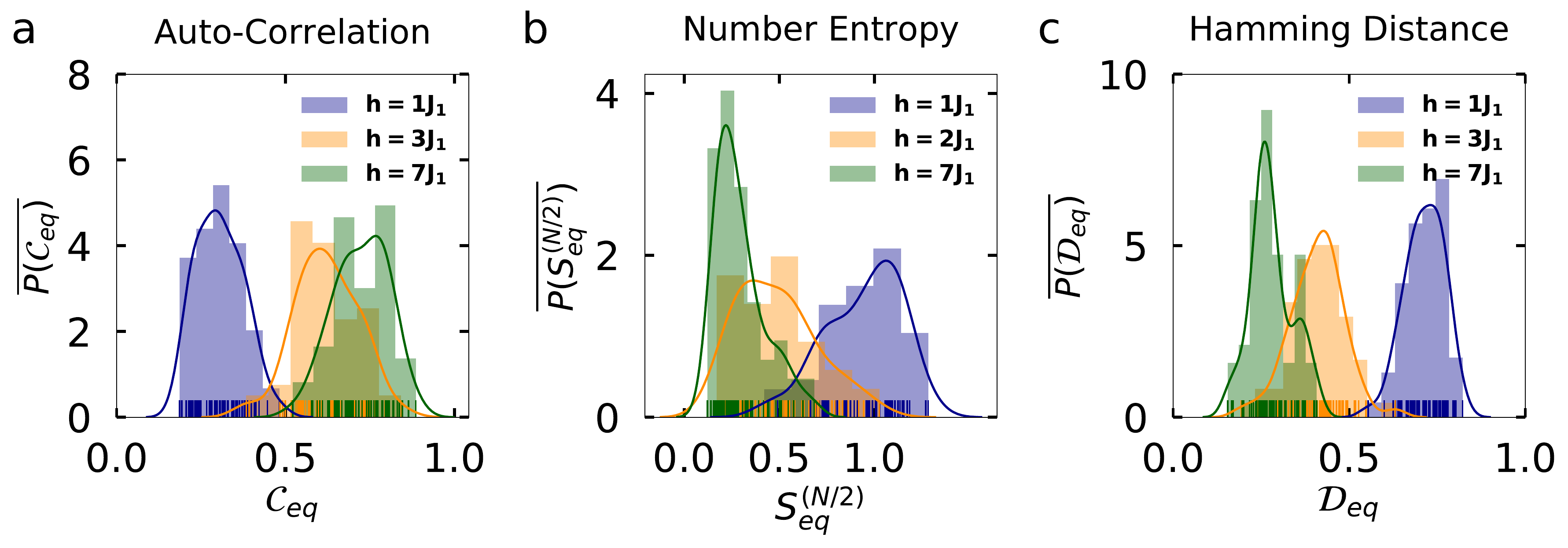}
		\caption{\textbf{Probability distributions of the equilibrium quantities.} 
			By repeating the experiment for $R=60$ random realizations, a set of equilibrium outcomes are obtained,  for which we fit a probability distribution $P(Q_{eq}^{(\mathbf{s})})$. Averaging over all given initial states leads to an initial state independent probability distribution $\overline{P(Q_{eq})}$. By considering different disorder strengths $h/J_1=1$ (ergodic), $h/J_1=2$ and $h/J_1=3$ (near the MBL transition) and $h/J_1=7$ (MBL) we estimate the probability distribution of: \textbf{a}, the auto-correlation function $\overline{P(\mathcal{C}_{eq})}$; \textbf{b}, the number entropy of half chain $\overline{P(S_{eq}^{(N/2)})}$; and \textbf{c}, the Hamming distance $\overline{P(\mathcal{D}_{eq})}$. In all the figures, the small rugs are the experimental data, the bars are the histograms and the solid lines are the fitting distributions. In the ergodic regime, the distribution is narrow as the system thermalizes and the final outcomes are determined solely by the thermal state. By increasing the disorder strength $h/J_1$, the outcomes can take a larger range of values as the final results heavily depend on the random potential pattern. Consequently, the probability distribution gets wider. As we go into the MBL regime, the dynamics tends to get frozen, so the final outcomes are mainly determined by the initial state and weakly depend on the random potential pattern. Therefore, the distribution gets narrower again.}
		\label{fig:fig3}
	\end{figure*}
	
	\begin{figure*}
		\includegraphics[width=1.\linewidth]{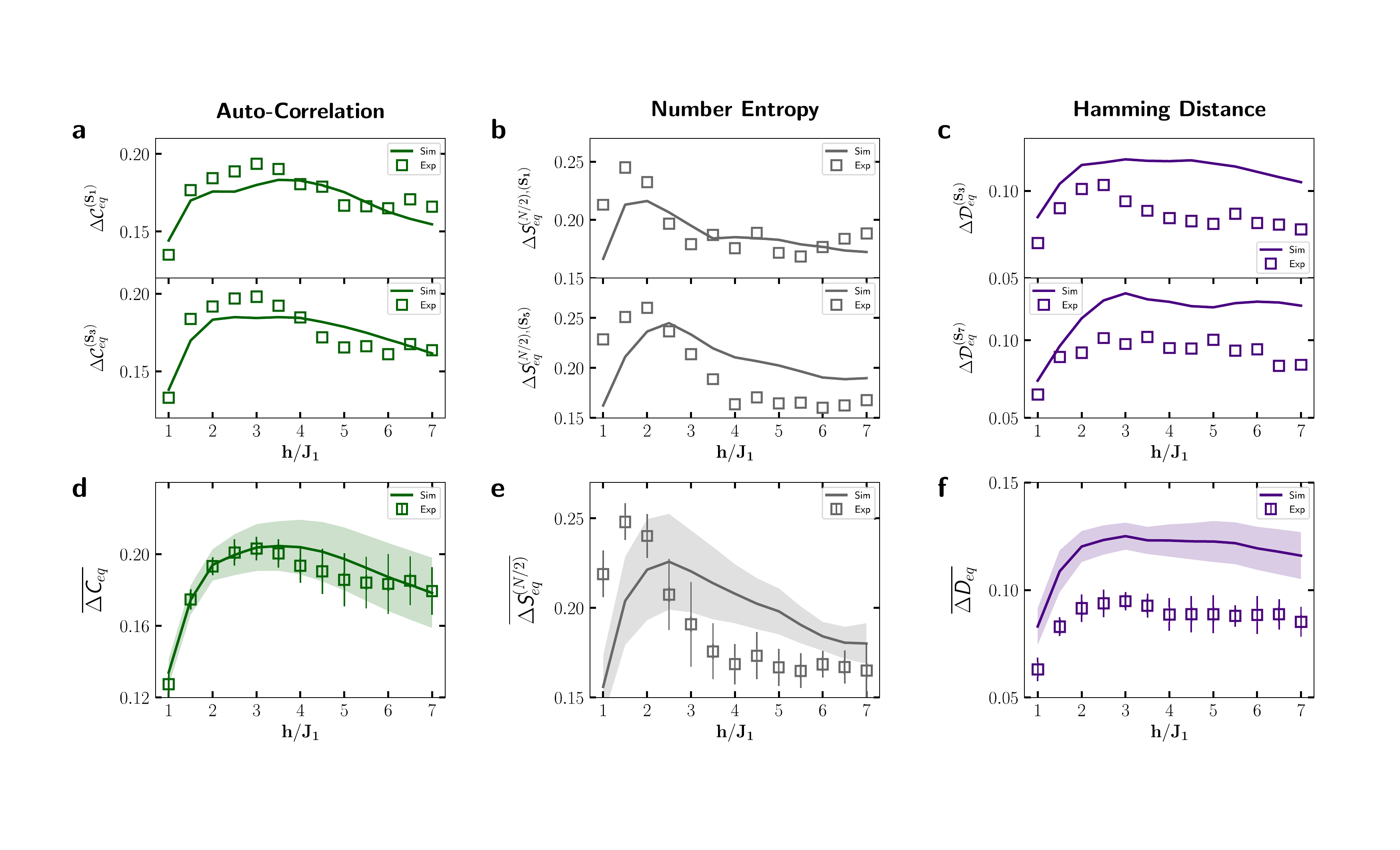}
		\caption{\textbf{Standard deviations of the quantities as a function of disorder strength.} The standard deviation $\Delta Q_{eq}^{(\textbf{s})}$ with respect to random samples for two representative initial states $|\textbf{s}_k\rangle$ (see the Supplementary Materials for other initial states) as a function of disorder strength $h/J_1$ for: \textbf{a}, the auto-correlation function $\Delta \mathcal{C}_{eq}^{(\textbf{s})}$; \textbf{b}, the number entropy of half chain $\Delta S_{eq}^{(N/2),(\textbf{s})}$; and \textbf{c}, the Hamming distance $\Delta \mathcal{D}^{(\textbf{s})}_{eq}$. 
			Remarkably, the $\Delta Q_{eq}^{(\textbf{s})}$ shows an evident peak around $h/J_1 \approx 3$, revealing the elusive MBL transition point. Moreover, the location of the peak slightly changes for each initial state which is a clear witness of the mobility edge. To present the average behavior with respect to the initial states we plot the average standard deviation $\overline{\Delta Q_{eq}}$ as a function of $h_1/J$ for: \textbf{d}, the auto-correlation function $\overline{\Delta \mathcal{C}_{eq}}$; \textbf{e}, the number entropy of half chain $\overline{\Delta S_{eq}^{(N/2)}}$; and \textbf{f}, the Hamming distance $\overline{\Delta \mathcal{D}_{eq}}$. The shadow around the theory simulations represents the behavior for various initial states.}
		\label{fig:fig4}
	\end{figure*}

\section*{Introduction}	
Ergodicity and thermalization principles are the foundations of statistical mechanics which imply that a many-body system forgets its local information as it evolves~\cite{rev1,thermal}. Strikingly, these principles can be violated when the thermalizing dynamics leads to the conservation of local information~\cite{rev5,integral}. Many-Body Localization (MBL) is the most recognized phenomenon for this remarkable feature~\cite{rev2,rev3,rev4,ergo}. 
The MBL is the reminiscent of Anderson localization~\cite{Anderson-Localization} when the particles of a disordered many-body system interact. 
The
most mysterious feature of the MBL physics is the transition point at which
an ergodic phase transforms into a localized one as the disorder strength increases. Despite several theoretical breakthroughs for characterizing the transition point~\cite{t1,EE1,Gap,Altman-Huse-MBL-Transition, Huse-Pal-MBL-Transition, Serbyn-Papic-Abanin-2015,negativity,Bayat-Memory-MBL}, its experimental observation is extremely challenging limited to low filling factors~\cite{spectrum} or weak interaction~\cite{greiner} regimes.\\
			
The MBL transition takes place as the strength of disorder exceeds a threshold value in comparison with the interaction coupling. Unlike quantum phase transition, which only affects the ground state, the MBL transition is more drastic and leaves its impact on the whole spectrum. This makes it difficult to investigate as, for instance, theoretically it can only be explored via exact diagonalization which restricts us to short chains~\cite{exact}. Interestingly, each energy eigenstate localizes at a different disorder strength, with the mid-spectrum eigenvectors requiring the maximum disorder. This phenomena, schematically shown in Fig.~\ref{fig:fig1}\textbf{a}, is known as mobility edge which has been investigated theoretically~\cite{t1} and some of its features have been observed experimentally~\cite{Mobility,Mobility2}.
While ergodic and MBL phases are well explored through thermalization studies~\cite{thermal} and investigating local conservation laws~\cite{integral}, several interesting features emerge around the MBL phase transition point which are less understood. This includes  scaling properties~\cite{t1,Gap,negativity}, anomalous transport~\cite{transport}, critical slowing down~\cite{slow} and a change in entanglement behavior~\cite{EE1} and energy level statistics~\cite{level}. Hence, detection, characterization and understanding the MBL transition point is highly desirable for both fundamental and practical purposes. 
Most of the quantities, e.g. von Neumann entropy~\cite{EE1}, level spacing statistics~\cite{level}, Schmidt gap~\cite{Gap} and entanglement negativity~\cite{negativity},  which have been introduced for identifying the transition point are not experimentally friendly, demanding either costly state tomography protocols~\cite{super} or the full knowledge 	of the energy spectrum~\cite{t1}. 
Therefore, most of the experiments~\cite{ion,super,super1,Zha2020,kim,spectrum,greiner,coldatom2,coldatom3} are performed either in the ergodic phase or deep in the MBL regime, leaving the MBL transition point unexplored.

Here, we propose an experimentally implement a protocol for detecting the MBL transition point using a superconducting transmon qubit array with size $N=12$ qubits. We initialize the system in various product states and let it 	evolve under the action of its disordered Hamiltonian until it reaches local 	equilibrium. Then, we measure three  different  quantities, namely time auto-correlation, number entropy and Hamming distance, which capture different spatial resolutions varying from single site to a block of finite size and the entire system, respectively. To see the effect of mobility edge, various initial states with different overlap pattern with the eigenstates of the system are considered. Each quantity is measured for several random ensembles. While the averaged quantities vary smoothly across the phase diagram, their standard deviation with respect to different random ensembles peaks at the critical point, 	revealing the MBL transition.


\section*{Superconducting Qubit Array}

We demonstrate our experiment on an array of $N{=}12$ superconducting transmon qubits, described by Bose-Hubbard model $(\hbar{=}1)$
	\begin{eqnarray}
	\widehat{H}&=&\sum  \limits_{\ell=1}^{N-1} J_1^{(\ell)}(\widehat{a}_{\ell}^{\dag}\widehat{a
	}_{\ell+1}+\widehat{a}_{\ell}\widehat {a}_{\ell+1}^{\dag})\cr &+& 
	\sum \limits_{\ell=1}^{N-2} J_2^{(\ell)} (\widehat{a}_{\ell}^{\dag}\widehat{a
	}_{\ell+2}+\widehat{a}_{\ell}\widehat {a}_{\ell+2}^{\dag}) \cr
	&+&\sum\limits_{\ell=1}^{N}\left[ (\omega+h_{_{\ell}})\widehat{n}_{\ell}+
	\frac{U}{2}\widehat {n}_{\ell}(\widehat{n}_{\ell}-1)\right]
	\label{ham}
	\end{eqnarray}
where $\widehat{a}_{\ell }^{\dag }$ ($\widehat{a}_{\ell }$) are the bosonic 	creation (annihilation) operators at site $\ell $ and $\widehat{n}_{\ell }=	\widehat{a}_{\ell }^{\dag }\widehat{a}_{\ell }$ is the corresponding number 	operator. The capacitive dipole-dipole interaction leads to the 	nearest-neighbor hopping $J_1^{(\ell)}$, with the average value of $J_1{=}1/N\sum_{\ell} J_1^{(\ell)} {\simeq} 2\pi {\times} 11.5$ MHz, and the next nearest neighbor coupling $J_2^{(\ell)}$, with the average value of $J_2{=}1/N\sum_{\ell} J_2^{(\ell)} {\simeq} 2\pi {\times} 1.2$ MHz. In order to generate disorder in the potential of the system, the qubit frequencies are tuned by DC and pulse signals to be the sum of a 	constant central frequency $\omega$ and a random value $h_{_{\ell }}$ 	which is drawn from a uniform distribution $[-h,h]$ with $h$ being the disorder strength. The nonlinear on-site interaction $U\approx -22J_1$ represents the excess of energy needed for having more than one boson at each site. The schematic of the quantum simulator is shown in Fig.~\ref{fig:fig1}\textbf{b} and more details about the device and its parameters are given in the supplementary material~\cite{SM}. 
	
\section*{Experimental Protocol}
	
We initialize the system in $10$ different product states $ \left\vert \Psi _{\mathbf{s}}(0)\right\rangle =\left\vert \mathbf{s} 	\right\rangle =\left\vert s_{1},s_{2},\cdots ,s_{N}\right\rangle \,,$ (where $s_{\ell }=0,1$ represents the number of bosons at site $\ell$) such that the 	filling factor is $f=\sum_{\ell =1}^{N}\left\langle \widehat{n}_{\ell 	}\right\rangle /N=1/2$. The system evolves under the action of the
Hamiltonian as $\left\vert \Psi_{\mathbf{s}}(t)\right\rangle =e^{-\imath 
\widehat{H}t}\left\vert \Psi _{\mathbf{s}}(0)\right\rangle $ and then the population configuration is measured for all sites. The protocol is schematically shown in Fig.~\ref{fig:fig1}\textbf{c}.
We note that, the Hamiltonian commutes with the total number of particles, i.e. 
$\sum_{\ell =1}^{N}\widehat{n}_{\ell }$, resulting in the conservation of the number of boson during the evolution. Each measurement has been repeated 	for at least 3,000 times. The measurement outcomes which do not have the right filling factors are excluded (see \cite{SM} for detailed discussions). To ensure we observe the general behavior of the
disordered Hamiltonian, we also consider $R=60$ distinct random realizations	for each initial state. This is accompanied by numerical simulations in which the bosonic modes of each site are truncated to $ n_{\ell }^{\max }=3$. \\
	
We first consider the auto-correlation function which we adopt from Ref.~\cite{auto} and is defined as 
\begin{equation}
    \mathcal{C}(t)=\frac{1}{N}\sum_{\ell =1}^{N}\left( \left[ 2\left\langle \widehat{
		n_{\ell }}\right\rangle (t)-1\right] \left[ 2\left\langle \widehat{n_{\ell }}
	\right\rangle (0)-1\right]\right)
\end{equation}
In the ergodic phase, the local population thermalizes at long times reaching $\left\langle \widehat{n_{\ell }}\right\rangle {\sim} 1/2$ which results in $\mathcal{C}
{\sim} 0$. Deep in the MBL phase the evolution of the system is almost frozen, namely entanglement grows only logarithmically in time \cite{LogEE}. This means that $\left\langle \widehat{n_{\ell }}\right\rangle (t) {\sim} \left\langle \widehat{n_{\ell }}\right\rangle (0)$, which then leads to $\mathcal{C}{\sim} 1$.
	
To go beyond the single site resolution, it is highly desirable to investigate entanglement dynamics in the system. In practice evaluating a true measure of entanglement, such as the von Neumann entropy, is challenging as it demands full quantum state tomography. Here, we instead measure the number entropy~\cite{number,number1} for a block of size $m$\ (for $1\leq m\leq N/2$), which is defined as
\begin{equation}
  S^{(m)}(t)=-\sum\limits_{n}p_{n}\log (p_{n}),  
\end{equation}
where $p_{n}$ is the
probability of finding $n$ bosons in the block of size $m$. Since the system conserves the total number of bosons, the number entropy $S^{(m)}(t)$ is a lower bound for the von Neumann entropy~\cite{number1} and recently has attracted a lot of attention~\cite{number}.
While in the main text we focus on $m=N/2$, we provide more detailed analysis for other choices of $m$ in the Supplementary Materials \cite{SM}.	

Finally, we consider a global quantity, namely Hamming distance~\cite{HD}, which quantifies how different configurations emerge in the global wave-function of the
system making it distinct from the initial one. For a system initialized in $\left\vert \Psi _{\mathbf{s}
}(0)\right\rangle =\left\vert \mathbf{s}\right\rangle $, the Hamming distance is define as 
\begin{equation}
    \mathcal{D}(t)=\sum\limits_{\mathbf{s}^{\prime }}d(\mathbf{s}^{\prime },\mathbf{s})P(\mathbf{s}^{\prime },t)
\end{equation}	
where $\mathbf{s}^{\prime}$ is the configuration that one finds at the output  with probability $P(\mathbf{s}^{\prime },t)$ and $0\le d(\mathbf{s}^{\prime },\mathbf{s}) \le 1$ is the classical Hamming distance between
the two configurations $\mathbf{s}$ and $\mathbf{s}^{\prime }$ (i.e. the number of flips that converts $\mathbf{s}$ to $\mathbf{s}
	^{\prime }$ divided by $N$ for normalization). In
	the ergodic phase $\mathcal{D}(t)$ is expected to grow in time, ideally
	reaching $1$, due to superposition of multiple configurations in the
	wave-function of the system. In contrast, in the MBL phase the Hamming
	distance remains small as the dynamics is almost frozen and cannot generate many new configurations.
	The Hamming distance has already been employed to distinguish the ergodic
	and MBL phases experimentally \cite{ion}.

	Following the footsteps of previous experiments \cite{ion,super,Zha2020,kim,spectrum,greiner,coldatom2,coldatom3}, we first plot the three quantities, namely $\mathcal{C}(t), S^{(N/2)}(t)$ and $\mathcal{D}(t)$, as a function of time for one instance of random realization with different disorder strengths varying from ergodic to MBL phases in Figs.~\ref{fig:fig2}\textbf{a}-\textbf{c}. All the three quantities
	tend to equilibrate after a short transition time and we find good match
	between the experiments and our numerical simulations, showing that the unitary evolution under
	the action of the Hamiltonian in Eq.~(\ref{ham}) reasonably simulates the behavior
	of the real quantum device. 
	The scrambling nature of the dynamics in the
	ergodic phase (i.e. $h/J_1{=}1$ ) makes it very distinct from and an almost frozen evolution in the MBL phase (i.e. $h/J_1{=}7$). 
	In order to better characterize the difference between the ergodic and MBL
	phases, we focus on the equilibrium values. Let $Q^{(\mathbf{s}
		,r)}(t)$ represent any of the three quantities which depends on time $t$,
	a single random instance $r$  and the initial state $\mathbf{s}$. We define the
	equilibrium value as $Q_{eq}^{(\mathbf{s},r)}=\frac{1}{M}\sum\limits_{i=1}^{M}Q^{(\mathbf{s},r)}(t_{i})$, where $t_{i}$'s are the
	measured times in the equilibrium regime and we specifically choose $M=5$ points in range $7.9\le Jt\le10.8$. To have a statistical analysis of the behavior at the equilibrium
	with respect to random realizations for a given disorder strength $h$, we define
	the ensemble average $\left\langle Q_{eq}^{(\mathbf{s})}\right\rangle =\frac{1
	}{R}\sum\limits_{r=1}^{R}Q_{eq}^{(\mathbf{s},r)}\,$\ and its corresponding
	standard deviation $\Delta Q_{eq}^{(\mathbf{s})}=\frac{1}{R^{1/2}}\sqrt{
		\sum\limits_{r=1}^{R}\left[ (Q_{eq}^{(\mathbf{s},r)}-\left\langle Q_{eq}^{(
			\mathbf{s})}\right\rangle \right] ^{2}}$, where $R=60$ is the total number
	of random realizations. In order to see the behavior of the system independent
	of the choice of $\mathbf{s}$, one can average over different initial states to
	get $\overline{\left\langle Q_{eq}\right\rangle }=\frac{1}{I}\sum\limits_{
		\mathbf{s}}\left\langle Q_{eq}^{(\mathbf{s})}\right\rangle $ and $\overline{
		\Delta Q_{eq}}=$ $\frac{1}{I}\sum\limits_{\mathbf{s}}\Delta Q_{eq}^{(
		\mathbf{s})}$, where $I=10$ is the total number of initial states (see \cite{SM} for the exact choices of the initial states). 
	In Figs.~\ref{fig:fig2}\textbf{d}-\textbf{f}, we plot the average quantities $\overline{\left\langle \mathcal{
			C}_{eq}\right\rangle },\overline{\left\langle S_{eq}^{(N/2)}\right
		\rangle }$ and $\overline{\left\langle \mathcal{D}_{eq}\right\rangle }$ as a
	function of the disorder strength $h/J_1$. All the quantities clearly show a
	transition from ergodic to MBL as $h/J_1$ increase. Although, none of these
	quantities can directly reveal the transition point from the ergodic to the
	MBL phase as all of them vary smoothly throughout the phase diagram. However, as we will see later, it is indeed the averaged standard deviation $\overline{
		\Delta Q_{eq}}$ which is the crucial quantity to characterize the MBL transition
	point.

It is often more insightful to investigate a probability distribution directly rather than its average value. In our experiment, thanks to fairly large 	number of random samples (i.e. $R=60$), one can estimate the probability
	distribution of $P(Q_{eq}^{(\mathbf{s})})$ at each disorder strength $h$.
	To be initial state independent, one can also average over initial states to obtain 
	$\overline{P(Q_{eq})}=\frac{1}{I}\sum\limits_{
		\mathbf{s}} P(Q_{eq}^{(\mathbf{s})}) $. In Figs.~\ref{fig:fig3}\textbf{a}-\textbf{c}, the probability distributions $\overline{P(\mathcal{C}_{eq})},\overline{P(S_{eq}^{(N/2)})}$ and $\overline{P(\mathcal{D}_{eq})}$ are
	depicted for three different disorder strengths. We notice that, in the ergodic
	regime the shape of the distribution is almost Gaussian with a small
	variance. This is due to the scrambling dynamics in the ergodic regime which
	thermalizes the system locally making it indistinguishable for different
	random instances (i.e. absence of memory). As the disorder strength increases the
	probability distribution deviates from Gaussianity and gets wider due to strong dependence
	on the random ensembles. In fact, near the MBL transition point ($h/J_1 \sim 3$)
	an interplay between thermalization of the ergodic phase and localization of the MBL regime makes the system very sensitive to the random instances resulting in a wide distribution. By further increasing the disorder the system enters the MBL regime where the dynamics is almost frozen (i.e. emergence of memory)
	and local subsystems remain close to their initial value for all random instances. This
	makes the role of each random instance irrelevant and thus the distribution gets narrower again. In Fig.~\ref{fig:fig1}\textbf{d}, this phenomenon is explained schematically.
	
	Inspired by the probability distribution in Figs.~\ref{fig:fig3}\textbf{a}-\textbf{c}, one can
	exploit the width of the distribution for each initial state $\left\vert \mathbf{s}
	\right\rangle$, quantified through the standard deviation $\Delta Q_{eq}^{(\textbf{s})}$, to infer the elusive MBL
	transition point. In {Figs.~\ref{fig:fig4}\textbf{a}-\textbf{c}}, we plot $\Delta \mathcal{C}_{eq}^{(\textbf{s})},\Delta S_{eq}^{(N/2),(\textbf{s})}$ and $\Delta \mathcal{D}_{eq}^{(\textbf{s})}$, each for two different initial states $|\textbf{s}\rangle$, as a function of the disorder strength $h/J_1$. Strikingly, $\Delta Q_{eq}^{(\textbf{s})}$ shows a
	very clear peak for all the three quantities, identifying the MBL transition point. Furthermore, each initial state peaks at a slightly different disorder strength $h/J_1$. This is a clear evidence of the mobility edge as each initial state has a different overlap pattern with the eigenstates of the Hamiltonian and thus localizes at a different $h/J_1$. 
	One can get 10 different transition points corresponding to $I=10$ initial states for each of the three quantities (see \cite{SM} for the exact values). It is well-known that for any given length, due to the finite size effect and the different convergence rates, distinct quantities may result in slightly different values of the MBL transition point~\cite{Gap}. To extract this point reliably, we adopt  a data analysis procedure, based on Bayesian inference (see \cite{SM} for details). The transition point is, thus, identified to be $h_{c}/J_1=2.7$ 
	from the experimental data and $h_{c}/J_1=3.3$ from the numerical simulations, with standard deviations determined as 1.1 and 1.2, respectively. The uncertainty comes from the limited number of random samples. The two values of transition points are not only in excellent agreement but also fully consistent with a pure theoretical investigation, based on the von Neumann entropy~\cite{Vn,Vn1}, with $h_{c}/J_1 {\sim} 2{-}3$ for $|U|/J_1 {\sim}20$. The average behavior, which are independent of the initial state, are shown in  {Figs.~\ref{fig:fig4}\textbf{d}-\textbf{f}} 
	as a function of $h/J_1$. All the three quantities peak at the transition point, matching with the theory prediction. 
	
\section*{Conclusion}
	
We have proposed and experimentally realized a protocol for detecting the elusive MBL transition point in a superconducting quantum simulator. The proposed protocol relies on the time evolution of a many-body system under the action of a disordered Hamiltonian. We have focused on the long time evolution where the system is expected to reach an equilibrium. 
Three quantities, namely auto-correlation, number entropy and Hamming distance, each with different spatial resolution, have been measured to estimate the MBL transition point. The protocol relies on the standard deviation of these quantities with respect to the random samples of the disordered potential. As our results show, across the phase diagram, the sensitivity to random samples is maximum at the MBL transition point resulting in a peak in the standard deviation. The MBL transition point, computed from the three different quantities, are fully consistent with each other and mach well with numerical simulations. In addition, evidence of mobility edge, represented by slightly different transition point for each initial state, can be observed. 
	
A remarkable point about our results is that, neither of the three quantities demands the costly quantum state tomography, needed for 	measuring  the von Neumann entropy~\cite{EE1} or the Schmidt gap~\cite{Gap}, thus can be easily extended to larger systems and those platforms which cannot perform tomography measurements.
In addition, our protocol is platform independent which provides a clear application for the Noisy Intermediate Scale Quantum (NISQ) simulators to shed light on a difficult problem in many-body physics. 	In future experiments, by changing the filling factors and developing adjustable couplers~\cite{Supremacy}, we can investigate larger area of the phase diagram and complete the observation of mobility edge for all energy scales.

\section*{ACKNOWLEDGMENTS}
We thank Q. Pan and Z.-H. Sun for helpful discussions. The authors thank the USTC Center for Micro- and Nanoscale Research and Fabrication for supporting the sample fabrication. The authors thank Shujuan Li for her help in sample fabrication. The authors also thank QuantumCTek Co., Ltd., for supporting the fabrication and the maintenance of room-temperature electronics. This research was supported by the National Key R\&D Program of China, Grant 2017YFA0304300 and No.2018YFA0306703, the Chinese Academy of Sciences, Anhui Initiative in Quantum Information Technologies, Technology Committee of Shanghai Municipality, National Science Foundation of China (Grants no. 11574380, no. 11905217 and no. 12050410253) and Natural Science Foundation of Shanghai (Grant no. 19ZR1462700). GDN thanks for the support from the China Postdoctoral Science Foundation for the grant number 2018M643436.

\widetext
\clearpage	

\renewcommand{\theequation}{S\arabic{equation}}
\renewcommand{\thefigure}{S\arabic{figure}}
\renewcommand{\bibnumfmt}[1]{[S#1]}
\renewcommand{\citenumfont}[1]{S#1}
\renewcommand{\thetable}{S\arabic{table}}

\setcounter{equation}{0}
\setcounter{figure}{0}
\setcounter{table}{0}
\setcounter{page}{1}

		
\begin{center}
	\textbf{\Large Supplementary Materials}
\end{center}


		\section{Superconducting quantum processor} 
		The experiment is performed on a 12-qubit superconduting quantum processor \cite{Zha2020}, arranged in a 1D array, as illustrated in Fig. 1\textbf{b} of the main text. The qubits are Xmon variant \cite{Barends2013} of transmon qubits \cite{Koch2007}.  Each nearest-neighboring qubit pairs are coupled via the capacitance between them, yielding an average nearest-neighbor coupling strength of $J_1/2\pi\simeq 11.5$ MHz and an average next nearest neighbor coupling of $J_2/2\pi\simeq 1.2$ MHz. 
		For each qubit, there are an inductively coupled flux ($Z$) and a capacitively coupled microwave ($XY$) control lines to realize state manipulation.
		Twelve individual resonators are dispersively coupled to the qubits to realize state readout. The resonators are divided into two groups and each group couples to a transmission line. We use frequency multiplexing technology to simultaneously readout all qubits' states. At the outside of each transmission line, an impedance-matched parametric amplifier (IMPA) is used to enhance the readout signal strength. 
		The parameters of the device is listed in Table.~\ref{tab:pef}. The averaged energy relaxation time $T_1$ and dephasing time $T_2^*$  are 51.9 $\mu$s and 8.1 $\mu$s, respectively. \\
		
		\begin{table*}[h]
			\centering
			\begin{tabular}{lllllllllllll}
				\hline
				\hline
				& $Q_1$ & $Q_2$ & $Q_3$ & $Q_4$ & $Q_5$ & $Q_6$ & $Q_7$ & $Q_8$ & $Q_9$ & $Q_{10}$ & $Q_{11}$ & $Q_{12}$ \\
				\hline
				$f_{01}/2\pi$ (GHz)          & 3.981 & 4.523 & 4.063 & 4.621 & 4.211 & 4.654 & 4.02  & 4.467 & 3.99  & 4.432    & 3.96     & 4.575    \\
				$f_{ah}/2\pi$ (MHz)          & -248  & -256  & -264  & -216  & -256  & -248  & -240  & -256  & -256  & -248     & -248     & -240     \\
				$T_1$ ($\mu$s)          & 47.6  & 44.8  & 68.8  & 51.8  & 40.7  & 33.3  & 62.5  & 63.3  & 70.5  & 56.5     & 43.4     & 39.8     \\
				$T_2^*$ ($\mu$s)        & 2.6   & 9.9   & 2.3   & 5.4   & 3.4   & 16.2  & 4.3   & 26.9  & 2.3   & 5.3      & 2.5      & 15.6     \\
				$f_r/2\pi$ (GHz)             & 6.450 & 6.477 & 6.506 & 6.537 & 6.570 & 6.601 & 6.633 & 6.654 & 6.684 & 6.716    & 6.742    & 6.769    \\
				$f_{00}$ (\%)           & 94.4  & 96.6  & 96.1  & 94.7  & 97.6  & 93.7  & 97.2  & 95.3  & 92.0  & 98.0     & 95.6     & 98.0     \\
				$f_{11}$ (\%)           & 88.6  & 89.2  & 89.1  & 89.8  & 90.8  & 88.5  & 89.6  & 90.4  & 82.7  & 93.4     & 87.5     & 92.2     \\
				IMPA gain (dB) & 16.1 & 13.3 & 12.9 & 13.5 & 13.0 & 14.2 & 14.2 & 14.4 & 14.5 & 14.8 & 14.1 & 15.1 \\
				\hline
				\hline
			\end{tabular}
			\caption{\textbf{Performance of the qubits.} $f_{01}$ and $f_{ah}$ are the idle frequencies and anharmonicities, respectively. $T_1$ and $T_2^*$ are the energy relaxation time and dephasing time, respectively. $f_r$ is the frequency of the corresponding readout resonator. $f_{00}$ and $f_{11}$ are the probabilities of correctly reading out the qubits' state for $\ket{0}$ and $\ket{1}$, respectively. IMPA gain is measured as the ratio of the readout signal strengths when IMPAs are turned on and off.}
			\label{tab:pef}
		\end{table*}
		
		\section{Experimental realization}
		\begin{figure}[h!]
			\includegraphics[width=1\linewidth]{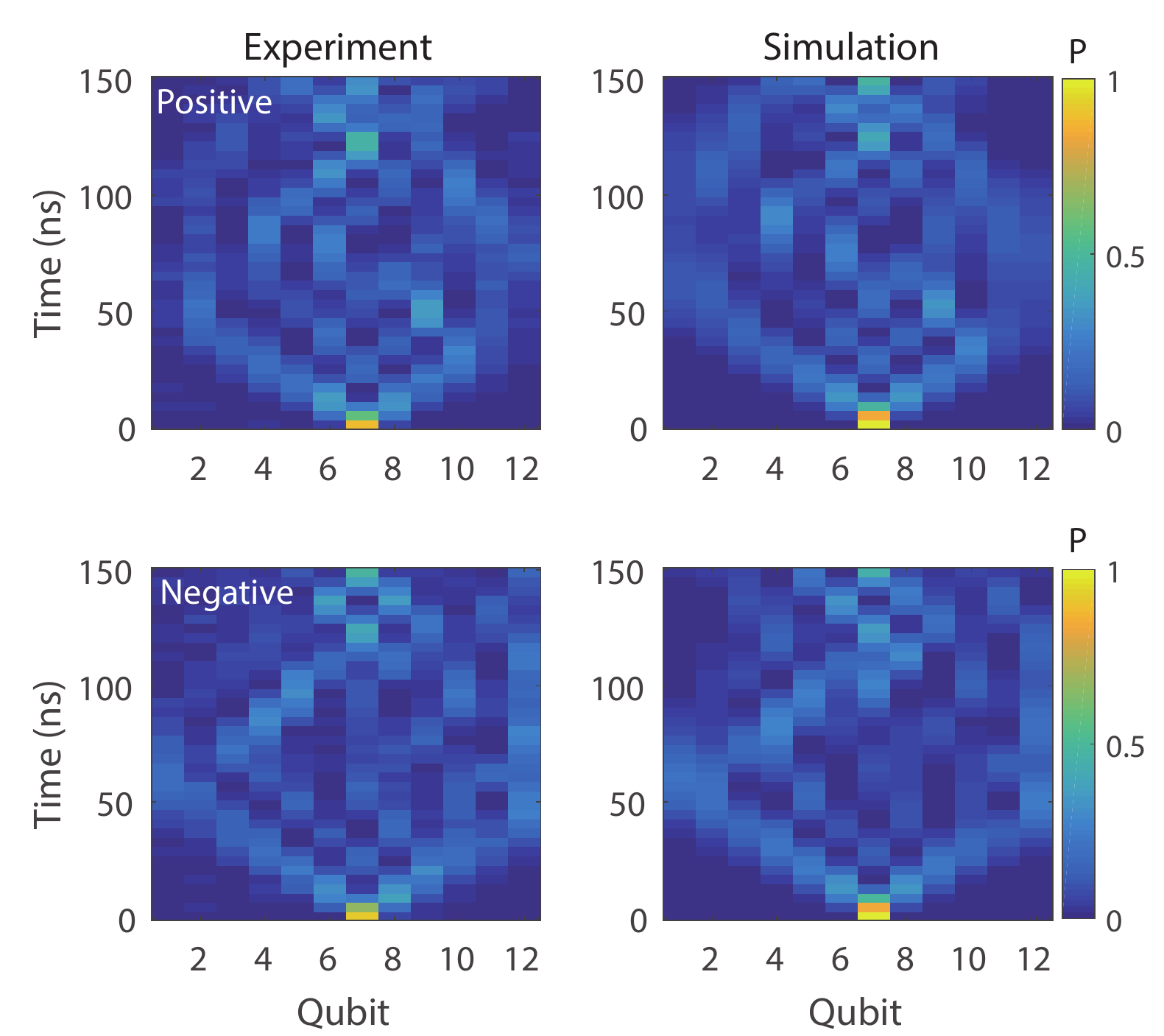}
			\caption{\textbf{Comparison of the experimental and simulated results in calibration of frequency alignment.} For positive case, the working frequencies of the twelve qubits ranges from 4.3225 GHz to 4.3775 GHz. For negative case, they are opposite. In both positive and negative cases, the $Q_7$ is excited to $\ket{1}$ and then start the evolution. The time evolutions for other qubits excited are not shown. The differences of the time evolution between the positive and negative cases come from the frequency differences on each sites. After optimizing the frequency differences, the maximum frequency difference is determined as 15.4 MHz on $Q_{11}$. }
			\label{fig:S1}
		\end{figure}
		As shown in Fig. 1\textbf{c} of the main text, the realization of this experiment consists of three steps: (i) state preparation; (ii) system evolution; and (iii) measurement. In state preparation, we apply $X$ gates to corresponding qubits to prepare the initial product state. After that, we detune all qubits to their working points, whose frequencies are randomized distributed in $[-h,h]$ in comparing with the central frequency 4.35 GHz. After the system evolves for a period $t$, we detune all qubits back to their idle frequencies and perform the simultaneous readout. In our experiment, we only perform the $\sigma_z$ projection measurements. We repeat each cycle for at least 3,000 times to obtain the statistical measurement results. The measurement outcomes which do not have the right filling factors are excluded. In the Supplementary Materials\texttt{} we discuss this in more details.  Between the cycles, all qubits are biased at their idle points for 300 $\mu$s to initialize all the qubits to their $\ket{0}$ states.
		
		In our experiments, we initially prepare the system in $10$ different product states $\left\vert \textbf{s} \right\rangle= \left\vert \textbf{s}_{k} \right\rangle$ ($k=1,...,10$), which are chosen to be:
		\protect\begin{eqnarray*}
			\left\vert \textbf{s} _{1}\right\rangle  &=&\left\vert
			0,1,1,0,1,0,1,1,1,0,0,0\right\rangle  \\
			\left\vert \textbf{s} _{2}\right\rangle  &=&\left\vert
			0,1,1,0,0,1,0,1,1,1,0,0\right\rangle  \\
			\left\vert \textbf{s} _{3}\right\rangle  &=&\left\vert
			0,1,0,1,1,1,0,1,1,0,0,0\right\rangle  \\
			\left\vert \textbf{s} _{4}\right\rangle  &=&\left\vert
			0,1,1,1,0,0,1,0,1,0,1,0\right\rangle  \\
			\left\vert \textbf{s} _{5}\right\rangle  &=&\left\vert
			0,1,1,1,1,0,0,1,1,0,0,0\right\rangle  \\
			\left\vert \textbf{s} _{6}\right\rangle  &=&\left\vert
			1,0,0,0,0,1,1,1,1,0,0,1\right\rangle  \\
			\left\vert \textbf{s} _{7}\right\rangle  &=&\left\vert
			1,0,0,0,1,1,1,0,0,1,1,0\right\rangle  \\
			\left\vert \textbf{s} _{8}\right\rangle  &=&\left\vert
			1,0,0,1,0,1,0,1,1,0,0,1\right\rangle  \\
			\left\vert \textbf{s} _{9}\right\rangle  &=&\left\vert
			1,0,0,1,1,0,1,1,0,0,1,0\right\rangle  \\
			\left\vert \textbf{s} _{10}\right\rangle  &=&\left\vert
			1,0,1,0,0,0,1,1,1,0,0,1\right\rangle \\
			\protect\end{eqnarray*}

		For each initial state, we consider $R=60$ distinct random realizations of disorders uniformly distributed in $[-h,h]$. We consider 13 disorder strengths ranging from $h=J_1$ to $h=7J_1$. In order to obtain the equilibrium value, we take $M=5$ different evolution times ranging from 110 ns to 150 ns. At each time, we measure the population configuration of the 12-qubit states and compute different quantities correspondingly.\\
		
		\section{Calibration of frequency alignment}
		The error of frequency alignment mostly comes from the residual nonlinear $Z$ cross talk~\cite{Yan2019}. In our experiment, the central frequency is set at $4.35$ GHz. Based on that, the working frequencies are randomized in range $[-h,h]$ in comparing with the central frequency. The calibration of the frequency alignment is thus important.
		The single round of the calibration consists of two steps. In the first step, we prepare the working frequencies in two different cases. For the positive case, the frequency difference between neighboring qubits is $+5$ MHz, and the working frequencies for $Q_1$ to $Q_{12}$ range from $4.3225$ GHz to $4.3775$ GHz. For the negative case, the frequency difference is $-5$ MHz. For these two cases, we firstly excite one qubit and then detune all qubits to their working frequencies for system evolution. After that, we detune them back to their idle frequencies for measurement. The population propagation is then measured as a function of time. For each case, we excite the twelve qubits sequentially, and then obtain the corresponding time-dependent population distributions. 
		In the second step, we run a Nelder-Mead optimization to find out the best estimation of the frequency differences. We assume that due to the residual $Z$ cross-talk, the frequency differences of each qubit site for the positive and negative cases are the same. Based on that, by numerically simulating the time evolution of the 12-qubit system, we obtain the population distributions as a function of time. We use the square sum of the differences between simulations and experiments as a cost function for optimization. An example of the comparison between the experimental and simulation results is shown in Fig.~\ref{fig:S1}, in which the different behavior for positive and negative cases, caused by the frequency differences, is presented. In the end of this round of calibration, we correct the frequency alignment by adding the differences to the central frequencies. 
		After several rounds of calibration, the maximum frequency difference is reduced from $15.4$ MHz to below $4.9$ MHz, smaller than 0.43$J_1/2\pi$. \\

		\section{Initial states}
		In this section, we show that the behavior of the quantities that we consider in the main text is general for various initial states. However, each initial state has a different overlap pattern with the eigenstates of the Hamiltonian. Each eigenstate localizes at a different disorder strength depending on its energy eigenvalue. Therefore, each initial state may show a different localization point. This is indeed an evidence for the presence of the mobility edge.

		\begin{figure*}[hbt!]
			\includegraphics[width=1\linewidth]{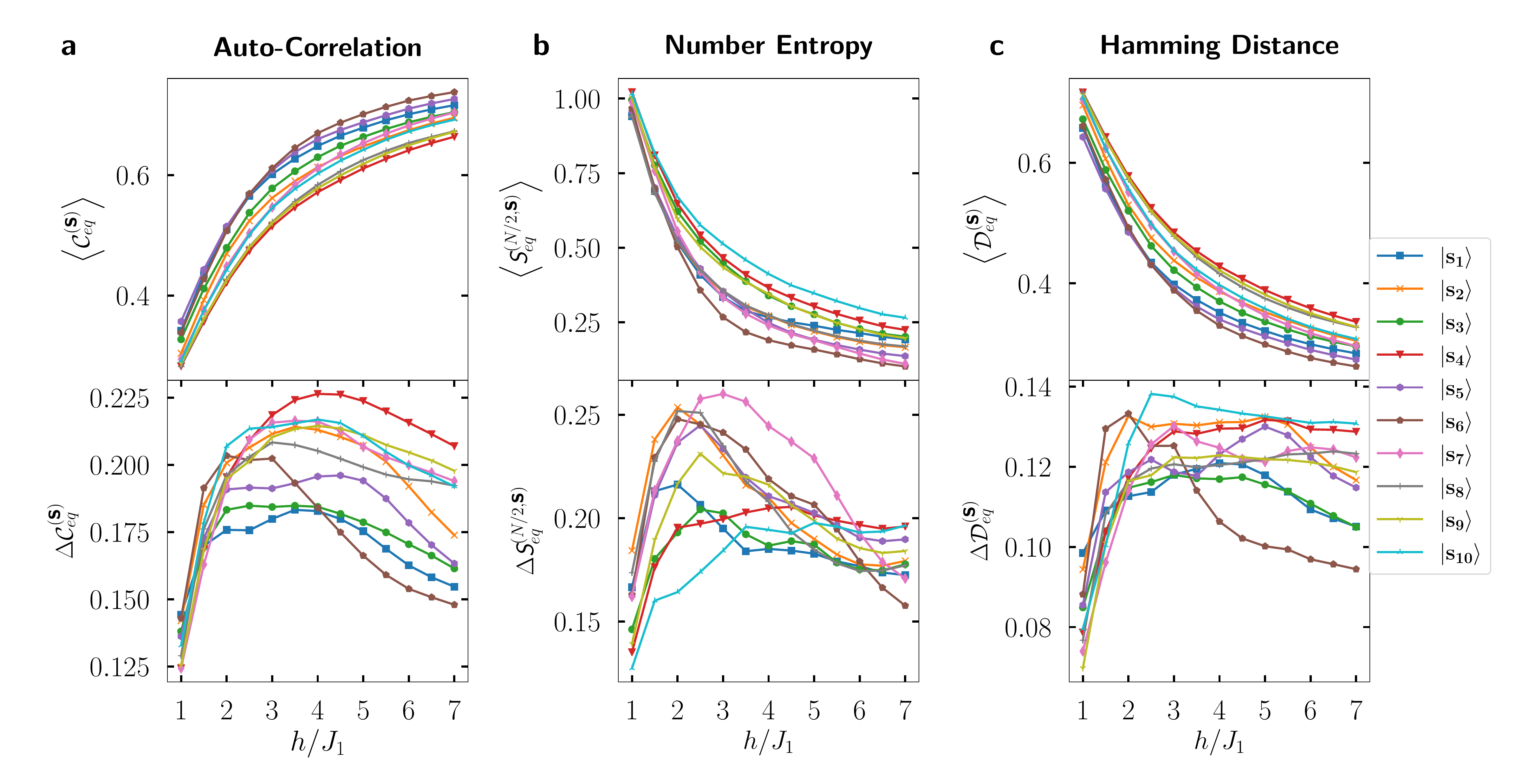}
			\caption{\textbf{Initial States Dependence }. The equilibrium values of the three quantities averaged over random realizations as a function of disorder strength $h/J_1$ for various initial states $\left\vert \textbf{s}\right\rangle=\left\vert \textbf{s}_k\right\rangle$ (upper panel). The panels respectively represent the mean value of: \textbf{a}, the auto-correlation $\left\langle \mathcal{C}_{eq}^{\textbf{(s)}}\right\rangle $; \textbf{b}, the number entropy $\left\langle \emph{S}_{eq}^{(N/2),\textbf{(s)}}\right\rangle $; and \textbf{c}, the Hamming distance $\left\langle \mathcal{D}_{eq}^{\textbf{(s)}}\right\rangle$. We note that, the three quantities smoothly change from their expected value in the ergodic phase to the MBL regime with similar qualitative behavior for all the initial states. The standard deviation of equilibrium values of the three quantities with respect to the $R=60$ random realizations as a function of disorder strength $h/J_1$ for various initial states $\left\vert \textbf{s}\right\rangle=\left\vert \textbf{s}_k\right\rangle$ is plotted in the lower panels. The lower panel respectively represents the standard deviation of: \textbf{a},the auto-correlation $\Delta \mathcal{C}_{eq}^{(\textbf{s})}$; \textbf{b}, the number entropy $\Delta \emph{S}_{eq}^{(N/2),(\textbf{s})}$; and \textbf{c}, the Hamming distance $\Delta \mathcal{D}_{eq}^{(\textbf{s})}$. We note that, for all the three quantities the peak of the standard deviation slightly varies for different initial states, witnessing the existence of the mobility edge.}

			\label{fig:S2}
		\end{figure*}
		\begin{table}[hbt!]
			\begin{tabular}{c|p{1.5cm}<{\centering}|p{1.5cm}<{\centering}|p{1.5cm}<{\centering}|p{1.5cm}<{\centering}|p{1.5cm}<{\centering}|p{1.5cm}<{\centering}}
				\hline
				\hline
				\multirow{2}{*}{}                    & \multicolumn{3}{c|}{Experiment} & \multicolumn{3}{c}{Simulation} \\ \cline{2-7} 
				&
				$ \mathcal{C}_{eq} $ &
				$ \emph{S}_{eq}^{(N/2)} $ &
				$ \mathcal{D}_{eq} $ &
				$ \mathcal{C}_{eq} $ &
				$ \emph{S}_{eq}^{(N/2)} $ &
				$ \mathcal{D}_{eq} $ \\ \hline
				$\left\vert \textbf{s} _1\right\rangle$      & 3.00   & 1.50  & 3.00    & 3.50   & 2.00 & 4.00 \\ \hline
				$\left\vert \textbf{s} _2\right\rangle$       & 3.00   & 2.00 & 6.00     & 3.50   & 2.00& 2.00  \\ \hline
				$\left\vert \textbf{s} _3\right\rangle$       & 3.00   & 1.50  & 2.50    & 2.50   & 2.50 & 3.00  \\ \hline
				$\left\vert \textbf{s} _4\right\rangle$      & 3.00   & 2.00  & 4.50   & 4.00   & 4.50  & 5.00\\ \hline
				$\left\vert \textbf{s} _5\right\rangle$      & 3.00   & 2.00  & 3.50   & 4.50   & 2.50  & 5.00\\ \hline
				$\left\vert \textbf{s} _6\right\rangle$      & 2.50   & 1.50  & 3.00   & 2.00   & 2.00  & 2.00\\ \hline
				$\left\vert \textbf{s} _7\right\rangle$       & 2.50   & 1.50 & 3.50    & 3.50   & 3.00 & 3.00 \\ \hline
				$\left\vert \textbf{s} _8\right\rangle$       & 3.00   & 1.50 & 3.50     & 3.00   & 2.00& 6.5  \\ \hline
				$\left\vert \textbf{s} _9\right\rangle$      & 3.00   & 1.50  & 5.00    & 4.00   & 2.50 & 4.00 \\ \hline
				$\left\vert \textbf{s} _{10}\right\rangle$   & 3.00   & 1.50 & 2.00     & 4.00   & 5.00& 2.5 \\ 
				\midrule[1pt]
				Mean                                 & 2.9  & 1.65  & 3.65     & 3.45   & 2.68  & 3.7 
				\\ \hline
				Standard deviation                   & 0.20  & 0.23  & 1.14    & 0.72   & 0.25  & 1.40  \\ \hline
				\hline
			\end{tabular}
			\caption{\textbf{The MBL Transition Point for Given Initial States.} The MBL transition point, extracted from the peak of the  $\Delta Q_{eq}^{(\textbf{s})}$, for the given initial states from both experimental data and numerical simulations. }
			\label{tab:1}
		\end{table}
		
		Here, we investigate the role of each initial state on  $\left\langle Q_{eq}^{\textbf{(s)}}\right\rangle $ and $\Delta Q_{eq}^{\textbf{(s)}}$ for the three quantities, namely the auto-correlation, the number entropy and Hamming distance. In order to ease the visualization, in the following, we only present the data from numerical simulation, since the experimental data follow the same behavior. 
		In Figs.~\ref{fig:S2}\textbf{a}-\textbf{c} (upper panel), 
		we plot the equilibrium auto-correlation  $\left\langle \mathcal{C}_{eq}^{\textbf{(s)}}\right\rangle $, the number entropy for a subsystem of half-chain $\left\langle \emph{S}_{eq}^{(N/2),\textbf{(s)}}\right\rangle $ and the Hamming distance $\left\langle \mathcal{D}_{eq}^{\textbf{(s)}}\right\rangle $ as a function of disorder strength for all the given initial states. As expected, all quantities smoothly change from the ergodic to the MBL phase. While, all the initial states qualitatively follow the same behavior there is small difference between them. To further evidence this, we plot the standard deviation $\Delta Q_{eq}^{\textbf{(s)}}$ as a function of disorder strength for all the given initial states $\left\vert \textbf{s} \right\rangle= \left\vert \textbf{s}_{k} \right\rangle$ in Figs.~\ref{fig:S2}\textbf{a}-\textbf{c} (lower panel). As it is clear from the figure, the peak of $\Delta Q_{eq}^{\textbf{(s)}}$ takes place at different disorder values for each initial state. This is a strong evidence for the observation of the mobility edge in our system.

		To have a better understanding, one may consider the time evolution of a given initial state $\vert  \textbf{s} \rangle$, which in general can be written as
		\begin{equation}
			\vert \Psi(t)\rangle=\sum_\alpha e^{-i\epsilon_\alpha t} \vert \epsilon_\alpha \rangle \langle \epsilon_\alpha | \textbf{s} \rangle
		\end{equation} 
		where $\epsilon_\alpha$ and $\vert \epsilon_\alpha \rangle$ are the eigenenergies and the eigenstates of the Hamiltonian $H$ for a particular random instance. The overlap of the quantum state of the system with each of the eigenstates $\vert \epsilon_\alpha \rangle$ is time independent and is given by $|\langle \epsilon_\alpha | \Psi(t) \rangle|^2=|\langle \epsilon_\alpha | \textbf{s} \rangle|^2$. Due to the mobility edge, for any given random instance $r$ each of the eigenstates $\vert \epsilon_\alpha \rangle$ localizes at a different disorder strength $h_c^{(\alpha,r)}$. Therefore, the transition point for the initial state $\vert  \textbf{s} \rangle$ and a random disorder instance $r$ is given by 
		\begin{equation}
			h_c^{(\textbf{s},r)}=\sum_{\alpha} |\langle \epsilon_\alpha | \textbf{s} \rangle|^2 h_c^{(\alpha,r)}.
		\end{equation}
		One can average over all random realizations as $h_c^{(\textbf{s})}=1/R \sum_{r}  h_c^{(\textbf{s},r)}$ to obtain the transition point $h_c^{(\textbf{s})}$ for the initial state $\vert  \textbf{s} \rangle$.
		In TABLE~\ref{tab:1}, we present the MBL transition point $h_c^{(\textbf{s})}$, determine as the location of the peak of $\Delta Q_{eq}^{\textbf{(s)}}$, for each initial state extracted from experimental and numerical simulation.

		\section{Data Processing and Post Selection of Measurement Results}
		\begin{figure*}[h!]
			\includegraphics[width=1\linewidth]{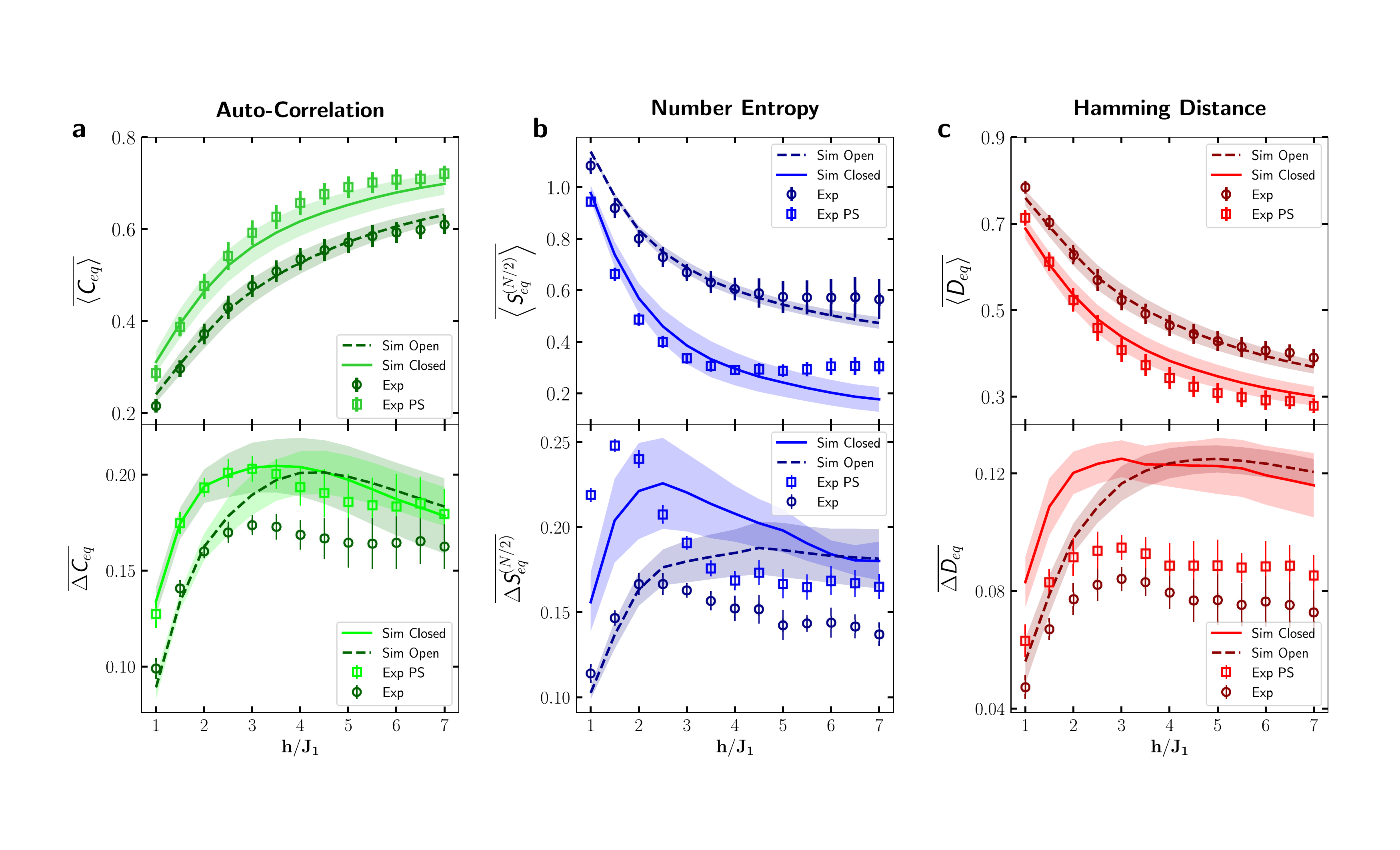}
			\caption{\textbf{Data Processing and Post Selection of Measurement Results.} In Figs.~\ref{fig:SM1}\textbf{a}-\textbf{c} (upper panel), 
				we plot the equilibrium auto-correlation  $\overline{\left\langle \mathcal{C}_{eq}\right\rangle} $, the half-chain number entropy $\overline{\left\langle \emph{S}_{eq}^{(N/2)}\right\rangle} $ and the Hamming distance $\overline{\left\langle \mathcal{D}_{eq}\right\rangle} $ as a function of disorder strength $h/J_1$ for the closed (Sim Closed) and open system simulations (Sim Open) as well as the raw (Exp) and post selected experimental (Exp PS) data. The error bars in the experimental data and shadows around the solid lines in the numerical simulations, represent the standard deviation with respect to random realizations. In Figs.~\ref{fig:SM1}\textbf{a}-\textbf{c} (lower panel), 
				we plot the standard deviation for the equilibrium auto-correlation  $\overline{ \Delta \mathcal{C}_{eq}} $, the half-chain number entropy $\overline{ \Delta \emph{S}_{eq}^{(N/2)}} $ and the Hamming distance $\overline{ \Delta \mathcal{D}_{eq}} $ as a function of disorder strength $h/J_1$ for the closed and open system simulations as well as the raw and post selected experimental data. We note that, for both the averaged values and the standard deviation, the post selected experimental data follows the simulations from the unitary evolution, while the raw data pursue the open quantum system evolution.}
			\label{fig:SM1}
		\end{figure*}
		The Bose-Hubbar Hamiltonian $H$, conserves the total number of excitations in the system. All the initial states that we have considered have the filling factor $f=1/2$. Nonetheless, due to experimental imperfections some of the measurement outcomes show a different filling factor which can be attributed to either the decay in transmonic qubits or imperfect readout. In addition, due to dephasing the dynamics may not be fully unitary, which can also induce errors in the results. In order to include these effects and possibly compensate them in our numerical simulations, we take two different approaches; (i) post selecting the experimental data, which means excluding the experimental measurement outcomes with wrong filling factors, and use unitary evolution for the numerical simulation; and (ii) keep the raw experimental data and instead use an open quantum system formulation for the numerical simulations. As mentioned before, the data shown in the main text is based on the first approach. Here, we provide a comparison between the two methods. For open quantum system evolution we use a Lindbladian master equation,
		\begin{equation}\label{sup_master}
			\frac{d\hat{\rho}}{dt}=-i\left[ \hat{H},\hat{\rho}\right] +\sum  \limits_{\ell=1}^{N} \left( \frac{\Gamma^{(\ell)}}{2}\mathcal{D}\left[\widehat{a}_{\ell}\right] \hat{\rho} 
			+\frac{\gamma^{(\ell)} }{2} \mathcal{D}\left[ \widehat{n}_{\ell }\right] \hat{\rho} \right).
		\end{equation}
		where $\hat{\rho}$ is the density matrix of the system, $\Gamma^{(\ell)}$ and $\gamma^{(\ell)}$ are the decay and dephasing rates of the qubit $\ell$, respectively. Moreover, the Lindblad term is denoted by 
		\begin{equation}\nonumber
			\mathcal{D}\left[ \hat{O}\right] =2\hat{O}\hat{\rho}\hat{O}^{\dagger }-\hat{%
				\rho}\hat{O}^{\dagger }\hat{O}-\hat{O}^{\dagger }\hat{O}\hat{\rho}.
		\end{equation}
		The values for decay and dephasing rates are taken from the TABLE I in the Methods section, such that $\Gamma^{(\ell)}=1/T_1^{(\ell)}$ and $\gamma^{(\ell)}=1/T_2^{*(\ell)}$. Since the open quantum system simulations of $N=12$ qubits, is computationally very costly we consider the average over $I=5$ initial states, in contrast to $I=10$ for experiments and unitary evolution. 
		In Figs.~\ref{fig:SM1}\textbf{a}-\textbf{c} (upper panel), 
		we plot the equilibrium auto-correlation  $\overline{\left\langle \mathcal{C}_{eq}\right\rangle} $, the half-chain number entropy $\overline{\left\langle \emph{S}_{eq}^{(N/2)}\right\rangle} $ and the Hamming distance $\overline{\left\langle \mathcal{D}_{eq}\right\rangle} $ as a function of disorder strength $h/J_1$ for the closed and open system simulations as well as the raw and post selected experimental data. Interestingly, the post selected data matches very well with the unitary simulations of the system. On the other hand, the raw experimental data matches with the open quantum system simulation, showing that indeed the decay and dephasing of the device during the time evolution are responsible for the deviation from the unitary dynamics. 
		Similarly, in Figs.~\ref{fig:SM1}\textbf{a}-\textbf{c} (lower panel), 
		we plot the standard deviation for the equilibrium auto-correlation  $\overline{\Delta \mathcal{C}_{eq}} $, the half-chain number entropy $\overline{\Delta \emph{S}_{eq}^{(N/2} }$ and the Hamming distance $\overline {\Delta \mathcal{D}_{eq}} $ as a function of disorder strength $h/J_1$ for the closed and open system simulations as well as the raw and post selected experimental data. Once again, we see that the post selected experimental data is qualitatively follow the results from the unitary evolution, while the raw data pursue the open quantum system evolution. Note that, the scale on the $y$-axis is small and thus the deviations between the curves are mainly visual.
		
		It is worth emphasizing that neither the post selection of experimental data nor the open quantum simulation can fully compensate the imperfections of the experiments. There are several reasons which contribute to this issue. First, the post selection of experimental data only takes into account the loss in the qubits and the first order readout errors (i.e. when only one qubit is flipped) while it cannot compromise the dephasing effects. Second, the Markovian model used for the open quantum system simulation in Eq.~(\ref{sup_master}) is very simplistic and ignores crosstalks and correlations between different qubits. Third, the experimental measurements of the device parameters, e.g. couplings, dephasing and decay rates etc., are inevitably prone to errors producing uncertainly in the parameter values used in the numerical simulations.
		
		\textbf{Numerical simulations.} All the numerical simulations were performed using the QUTIP Python toolbox~\cite{QuTIP}. In particular, for the time evolution we used the QUTIP mesolve master equation solver and the Hamiltonian parameters were taken from  Table~\ref{tab:pef}. The local bosonic Hilbert space is truncated at $n_{max}=3$. To see the accuracy of this truncation the commutators $\langle \Psi _{\mathbf{s}}(t)|[ \widehat{a}_{\ell },\widehat{a}_{\ell }^{\dag }]|\Psi _{\mathbf{s}}(t)\rangle=1\pm \epsilon$  were computed for all evolved states which showed $\epsilon\leq10^{-5}$. In fact, the good agreement between the theory and experiments in our data shows that the assumption of unitary evolution is valid.\\
		
		\section{Number entropy for different block sizes}
		\begin{figure*}[h!]
			\includegraphics[width=1\linewidth]{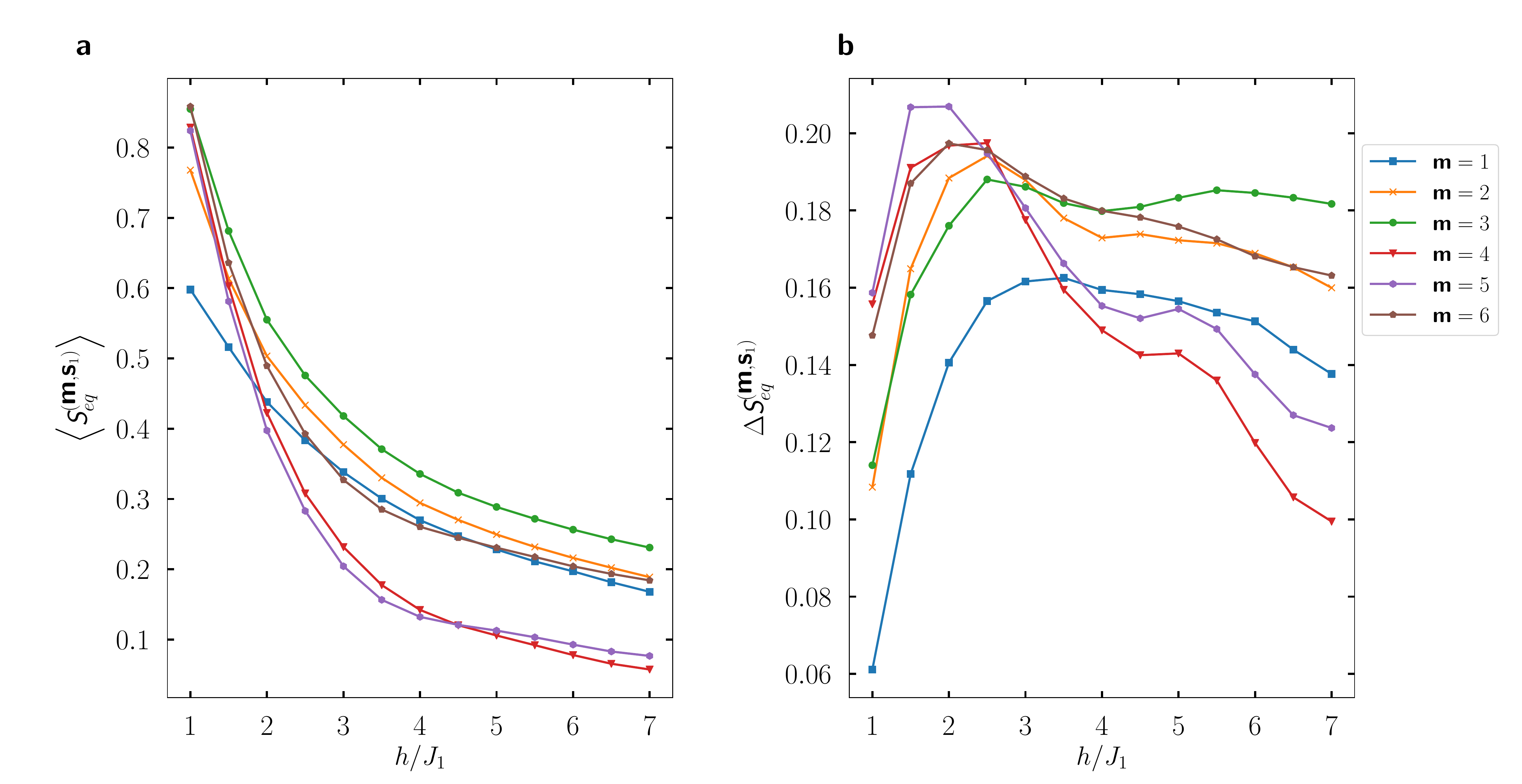}
			\caption{\textbf{Number Entropy of Different Subsystems}. \textbf{a}, The equilibrium number entropy $\left\langle \emph{S}_{eq}^{(m),(\textbf{s})}\right\rangle $ as a function of disorder strength $h/J_1$, for different subsystem size $m$ and the initial state $\vert \textbf{s}\rangle=\vert \textbf{s}_1\rangle$. As $m$ gets larger the maximum obtainable number entropy increases, which can be seems in the ergodic phase, where larger subsystems archive higher entropies. \textbf{b}, The standard deviation of the number entropy with respect to $R=60$ disorder samples $\Delta\emph{S}_{eq}^{(m),(\textbf{s})}$ as a function of disorder strength $h/J_1$ for different subsystem size $m$ and the initial state $\vert \textbf{s}\rangle=\vert \textbf{s}_1\rangle$. As the subsystem size $m$ increases the location of the peak mostly remains around $h/J_1 \sim 2-3$, indicating that the number entropy can characterize the MBL transition better when the subsystem is larger, namely $m \sim N/2$.
			}
			\label{fig:fig6}
		\end{figure*}
		The Bose-Hubbard model conserves the total number of excitations, that leads to a link between the particle numbers of the two complimentary subsystems. In the main text, we focused on the half-chain number entropy as a tool to characterize the MBL transition. To see the generality of the behavior of this quantity, in this section, we present the number entropy with different subsystem sizes. In Fig.~\ref{fig:fig6}\textbf{a}, we plot the equilibrium number entropy $\left\langle \emph{S}_{eq}^{(m),(\textbf{s})}\right\rangle $ as a function of disorder strength $h/J_1$, for different subsystem size $m$ and the initial state $\vert \textbf{s}\rangle=\vert \textbf{s}_1\rangle$. As the block size $m$ gets larger the number of possible outcomes for the excitation in the block increases and thus the achievable entropy becomes larger. This can be observed in Fig.~\ref{fig:fig6}\textbf{a}, when the system is in the ergodic regime, namely small $h/J_1$. In the MBL regime, since the dynamics is almost frozen the number entropy is always small for all subsystems sizes.
		In Fig.~\ref{fig:fig6}\textbf{b}, we display the standard deviation of the equilibrium number entropy $\Delta\emph{S}_{eq}^{(m),(\textbf{s})}$ as a function of disorder strength $h/J_1$ for different subsystem size $m$ and the initial state $\vert \textbf{s}\rangle=\vert \textbf{s}_1\rangle$.
		By increasing the system size $m$, the location of the peak of the standard deviation mostly takes place at around $h_/J_1 \sim 2-3$ showing that for detecting the MBL transition using larger subsystem sizes is more reliable.

		\begin{figure*}[hbt!]
			\includegraphics[width=1\linewidth]{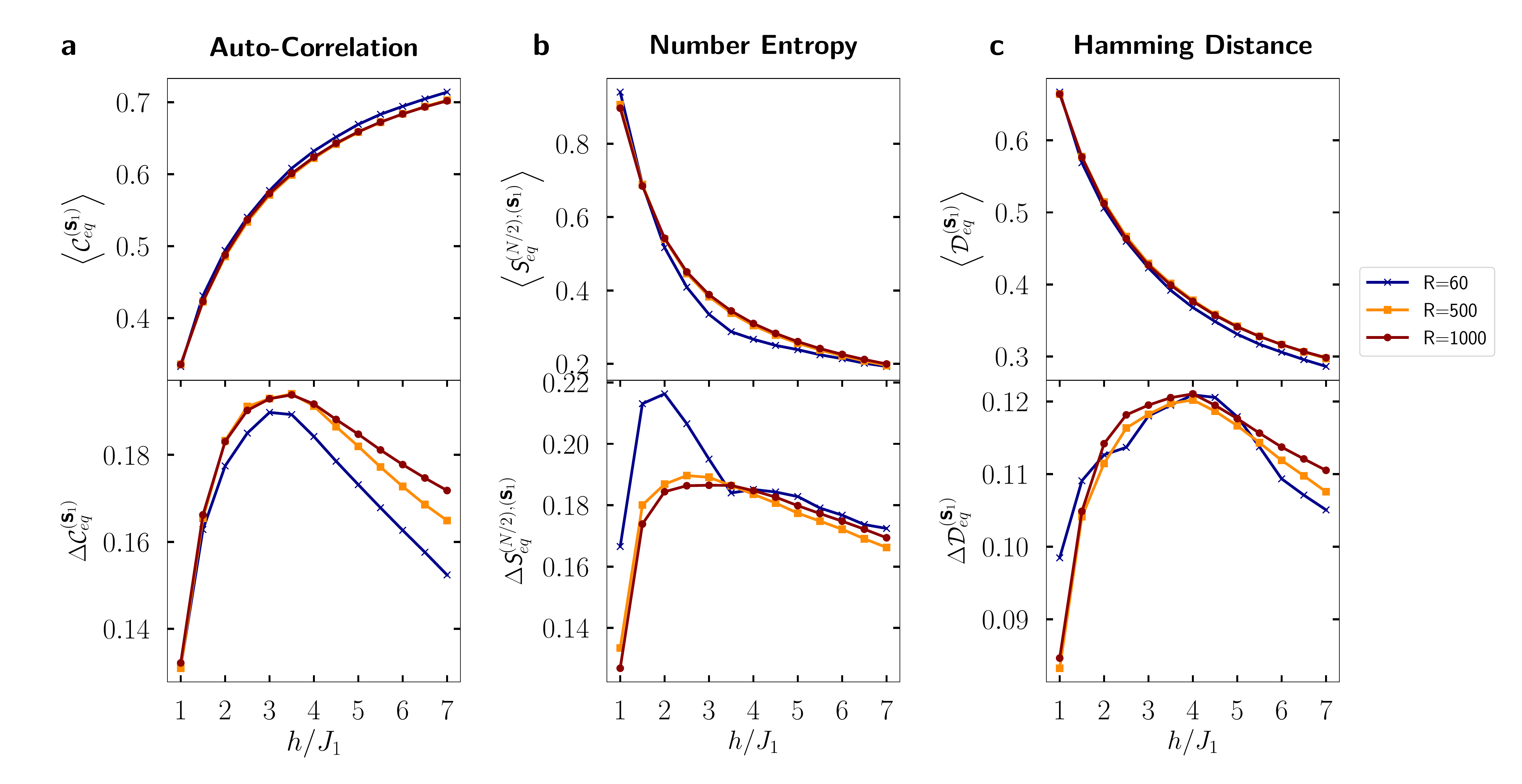}
			\caption{\textbf{Convergence with Respect to the Number of Random Realizations}. The three quantities are investigated for three different number of random realizations, namely $R=60$, $R=500$, and $R=1000$, computed for the initial state $\left\vert \textbf{s} _{1}\right\rangle$. 
				The upper panel represents the mean values of: \textbf{a}, the auto-correlation $\left\langle \mathcal{C}_{eq}^{(\textbf{s}_1)}\right\rangle $; \textbf{b}, the half-chain number entropy  $\left\langle \emph{S}_{eq}^{(N/2),(\textbf{s}_1)}\right\rangle $; and  \textbf{c}, the Hamming distance $\left\langle \mathcal{D}_{eq}^{(\textbf{s}_1)}\right\rangle $ as a function of $h/J_1$. In the lower panel the standard deviations are plotted as a function of $h/J_1$ for: \textbf{a}, the auto-correlation $\Delta\mathcal{C}_{eq}^{(\textbf{s}_1)} $; \textbf{b}, the half-chain number entropy $ \Delta \emph{S}_{eq}^{(N/2),(\textbf{s}_1)} $; and  \textbf{c}, the Hamming distance $ \Delta\mathcal{D}_{eq}^{(\textbf{s}_1)} $. The good agreement between the curves, specially the location at which the standard deviation peaks, shows that the choice of $R=60$ random realizations can faithfully captures the statistical behavior with respect to disorder in the system. }
			\label{fig:fig7}
		\end{figure*}

		\section{Convergence with respect to the number of random ensemble}
		
		As mentioned before, in our experiment, we employ $10$ different initial states and $R=60$ random samples. The role of different initial states has already been discussed and in this section we focus on the choice of the number of random realizations $R$. We compare the behavior of our quantities for the initial state $\left\vert \textbf{s} _{1}\right\rangle$ when the number of random ensembles is taken to be $R=60$ (as performed in our experiment), $R=500$ and $R=1000$ for which all quantities are expected to converge, . We find that, all the quantities are indeed converged even for $R=60$ random realizations. In Fig.~\ref{fig:fig7}\textbf{a}-\textbf{c} upper panel (lower panel) we plot, \textbf{a}, the auto-correlation $\left\langle \mathcal{C}_{eq}^{(\textbf{s}_1)}\right\rangle $ ($\Delta\mathcal{C}_{eq}^{(\textbf{s}_1)} $); \textbf{b}, the half-chain number entropy  $\left\langle \emph{S}_{eq}^{(N/2),(\textbf{s}_1)}\right\rangle $ ($ \Delta \emph{S}_{eq}^{(N/2),(\textbf{s}_1)} $); and  \textbf{c}, the Hamming distance $\left\langle \mathcal{D}_{eq}^{(\textbf{s}_1)}\right\rangle $ ($ \Delta\mathcal{D}_{eq}^{(\textbf{s}_1)} $) versus disorder strength $h/J_1$. Both the mean value and the standard deviation shows convergence with $R=60$ random realizations.

		\section{Data analysis based on Bayesian inference.} For each quantity measured in our experiment , we start with 10 different initial states. The 10 data points obtained from the peaks are within one groups. The three quantities are independent, and the groups for the three quantities are hierarchical. Therefore, to estimate the total mean, which corresponds to the transition point, we use a multilevel modeling approach \cite{Bayesian}, which allows us to consider the group level variance and individual level variance. 
		
		The data points $Y_{i,j}$, where $i=\mathcal{C},\mathcal{S},\mathcal{D}$, and $j=1,...,10$, are assumed to be independently normally distributed within each group $i$, 
		$Y_{i,j}|\theta_i\sim N(\theta_i,\delta^2),$
		where $\theta_i$ is the group mean and $\delta^2$ is the individual level variance. The group means $\theta_i$ are assumed to follow a normal distribution with mean $\mu$ and group level variance $\tau^2$, 
		$\theta_i\sim N(\mu,\tau^2).$
		
		Based on this model, we use Bayesian modeling with MCMC algorithms~\cite{Bayesian} to estimate group mean $\mu$. Gibbs sampling is used as the MCMC sampler~\cite{Bayesian}. For the model used in this study, the form of posterior distribution is already known, and based on that, the conditional posterior distributions for the three parameters are then generated. The start values are sampled from non-informative priors distributions. After assigning starting points, Gibbs sampler randomly draw samples from conditional posterior distributions. At a time, only one component of the parameters is updated. In this study, we have totally three individual chains. The converge is determined by all three chains when between-chain differences are small enough. For each chain, the total iterations is $1\times10^6$, and we discard first $8\times10^5$ points. Among the last $2\times10^5$ iterations, we retain the samples every 10 iterations to reduce auto-correlation. The final posterior probability distributions for each parameter is then determined with the retained samples. From the mode of the distributions, we obtain the $\mu$ and $\tau$, which corresponds to the estimation value of transition point and its 95\% confidence interval, respectively. 
		


\begin{thebibliography}{99}
		\bibitem{rev1} M. Rigol, V. Dunjko, and  M. Olshanii, Thermalization and its mechanism for generic isolated quantum systems,  \textit{Nature} \textbf{452}, 854 (2008).
	    \bibitem{thermal} J. Eisert, M. Friesdorf, and C. Gogolin, Quantum many-body systems out of equilibrium. \textit{Nat. Phys.} \textbf{11},124 (2015).
	    \bibitem{rev5} D. A. Abanin, E. Altman, I. Bloch, and M.  Serbyn, Colloquium: Many-body localization, thermalization, and entanglement, \textit{Rev. Mod. Phys.} \textbf{91}, 021001 (2019).
	    \bibitem{integral} M. Serbyn, Z. Papic, and D. A. Abanin, Local conservation laws and the structure of the many-body localized states, \textit{Phys. Rev. Lett.} \textbf{111}, 127201 (2013).
		\bibitem{rev2} R. Nandkishore, and D. A. Huse, Many-body localization and thermalization in quantum statistical mechanics, \textit{Annu. Rev. Condens. Matter Phys.} \textbf{6}, 15 (2015).
		
		 \bibitem{rev3} C. Gogolin, and J. Eisert, Equilibration, thermalisation, and the emergence of statistical mechanics in closed quantum systems, \textit{Rep. Prog. Phys.} \textbf{79}, 056001 (2016).
		 
		\bibitem{rev4} F. Alet, and N. Laflorencie,  Many-body localization: An introduction and selected topics, \textit{C. R. Phys.} \textbf{19}, 498 (2018).
		
		
		
	
		\bibitem{ergo} L. D’Alessio, Y. Kafri, A. Polkovnikov, and M. Rigol, From quantum chaos and eigenstate thermalization to statistical mechanics and thermodynamics, \textit{Adv. Phys.} \textbf{65}, 239–362 (2016).
		
	\bibitem{Anderson-Localization} P. W. Anderson, Absence of diffusion in certain random lattices, \textit{Phys. Rev.} \textbf{109}, 1492 (1958).
		
	\bibitem{Gap}  J. Gray, S. Bose, and A. Bayat, Many-body localization transition: Schmidt gap, entanglement length, and scaling, \textit{Phys. Rev. B} \textbf{97}, 201105 (2018).
		
	\bibitem{negativity} J. Gray, A. Bayat, A. Pal, and S. Bose, Scale Invariant Entanglement Negativity at the Many-Body Localization Transition, arXiv:1908.02761.
		
	\bibitem{Bayat-Memory-MBL} A. Nico-Katz, A. Bayat, and S. Bose, Information-Theoretic Memory Scaling in the Many-Body Localization Transition, arXiv:2009.04470.	
	
	\bibitem{Altman-Huse-MBL-Transition} R. Vosk, D. A. Huse, and E. Altman. Theory of the Many-Body Localization Transition in One-Dimensional Systems. \textit{Phys. Rev. X} \textbf{5}, 031032 (2015).
	
	\bibitem{Huse-Pal-MBL-Transition} A. Pal, and D. A. Huse, Many-body localization phase transition, \textit{Phys. Rev. B} \textbf{82}, 174411 (2010).
	
	
	\bibitem{Serbyn-Papic-Abanin-2015} M. Serbyn, Z. Papic, D. A. Abanin, Criterion for Many-Body Localization-Delocalization Phase Transition, \textit{Phys. Rev. X} \textbf{5}, 041047 (2015).
		\bibitem{t1} D. J. Luitz, N. Laflorencie, and F. Alet, Many-body localization edge in the random-field Heisenberg chain, \textit{Phys. Rev. B} \textbf{91}, 081103 (2015).
		
		
		\bibitem{EE1} V. Khemani, S. P. Lim, D. N. Sheng, and  D. A. Huse,  Critical properties of the many-body localization transition, \textit{Phys. Rev. X} \textbf{7}, 021013 (2017).
		
		
		\bibitem{spectrum} P. Roushan \textit{et al.}, Spectroscopic signatures of localization with interacting photons in superconducting qubits, \textit{Science} \textbf{358}, 1175 (2017).
		
		\bibitem{greiner} M. Rispoli, A. Lukin, R. Schittko, S. Kim, M. E. Tai, J. Leonard, and M. Greiner, Quantum Critical Behaviour at the Many-Body Localization Transition, \textit{Nature} \textbf{573}, 385 (2019).
		
		
		
		\bibitem{exact} F. Pietracaprina, N. Mace, D. J. Luitz, and F. Alet, Shift-invert diagonalization of large many-body localizing spin chains, \textit{SciPost Phys.} \textbf{5}, 045 (2018).
		
		\bibitem{Mobility}  T. Kohlert, S. Scherg, X. Li, H. P. Lüschen, S. D. Sarma, I. Bloch, and M. Aidelsburger, Observation of many-body localization in a one-dimensional system with a single particle mobility edge. \textit{Phys. Rev. Lett.} \textbf{122}, 170403 (2019).
		
		\bibitem{Mobility2} Q. Guo, C. Cheng, Z.-H. Sun, Z. Song, H. Li, Z. Wang, W. Ren, H. Dong, D. Zheng,
Y.-R. Zhang, R. Mondaini, H. Fan, H. Wang, Observation of energy resolved many-body localization. \textit{Nat. Phys.}, 1 (2020).
		
	
		
		
		
		
		
	
		
		\bibitem{transport}  A. C. Potter, R. Vasseur, and S. A. Parameswaran, Universal properties of
		many-body delocalization transitions. \textit{Phys. Rev. X} \textbf{5}, 031033 (2015).
		
		\bibitem{slow} P. Bordia, H. Lüschen, S. Scherg, S. Gopalakrishnan, M. Knap, U. Schneider, and I. Bloch, Probing slow relaxation and many-body localization in
		two-dimensional quasiperiodic systems, \textit{Phys. Rev. X} \textbf{7}, 041047 (2017).
		
		\bibitem{level} V. Oganesyan, and  D. A. Huse, Localization of interacting fermions at high temperature, \textit{Phys. Rev. B} \textbf{75}, 155111 (2007).
		
		
		
		\bibitem{super} K. Xu, J.-J. Chen, Y. Zeng, Y.-R. Zhang, C. Song, W. Liu, Q. Guo, P. Zhang, D. Xu, H. Deng, K. Huang, H. Wang, X. Zhu, D. Zheng, and H. Fan, Emulating many-body localization with a superconducting quantum processor. \textit{Phys. Rev. Lett.} \textbf{120}, 050507 (2018).
		
		\bibitem{super1} Q. Guo \textit{et al.}, Stark many-body localization on a superconducting quantum processor. arXiv:2011.13895, (2020).
		\bibitem{ion} J. Smith, A. Lee, P. Richerme, B. Neyenhuis, P. W. Hess, P. Hauke, M. Heyl, D. A. Huse, and C. Monroe, Many-body localization in a quantum simulator with programmable random disorder. \textit{Nat. Phys.} \textbf{12}, 907 (2016).
		
		
		\bibitem{Zha2020} C. Zha \textit{et al.}, Ergodic-localized junctions in a periodically-driven spin chain. \textit{Phys. Rev. Lett.} \textbf{125}, 170503 (2020).
		
		\bibitem{kim} A. Smith, M. S. Kim,  F. Pollmann, and J. Knolle, Simulating quantum many-body dynamics on a current digital quantum computer. \textit{npj Quant. Inform.} \textbf{5}, 106 (2019).
		
		
		
		\bibitem{coldatom2} J.-y. Choi, S. Hild, J. Zeiher, P. Schauß, A. Rubio-Abadal, T. Yefsah, V. Khemani, D. A. Huse, I. Bloch, and C. Gross, Exploring the many-body localization transition in two dimensions, \textit{Science} \textbf{352}, 1547 (2016).
		
		\bibitem{coldatom3} H. P. Lüschen, P. Bordia, S. Scherg, F. Alet, E. Altman, U. Schneider, and I. Bloch, Observation of slow dynamics near the many-body localization transition in one-dimensional quasiperiodic systems. \textit{Phys. Rev. Lett.} \textbf{119}, 260401 (2017).
		
		\bibitem{SM} See Supplemental Materials for more details.
		
		
		\bibitem{auto} S. Iyer, V. Oganesyan, G. Refael, and D. A. Huse, Many-body localization in a quasiperiodic system, \textit{Phys. Rev. B} \textbf{87},  134202 (2013).
		
		\bibitem{LogEE} J. H. Bardarson, F. Pollmann, and J. E. Moore,  Unbounded growth of entanglement in models of many-body localization, \textit{Phys. Rev. Lett.} \textbf{109}, 017202 (2012). 
		
		\bibitem{number} A. Lukin, M. Rispoli, R. Schittko, M. E. Tai, A. M. Kaufman, S. Choi, V. Khemani, J. Léonard, and M. Greine, Probing entanglement in a many-body–localized system, \textit{Science} \textbf{364}, 256–260 (2019).
		
		\bibitem{number1}  M. Kiefer-Emmanouilidis, R. Unanyan,  M. Fleischhauer, and J. Sirker, Evidence for unbounded growth of the number entropy in many-body localized phases, \textit{Phys. Rev. Lett.} \textbf{124}, 243601 (2020).
		
		\bibitem{HD} P. Hauke, and M. Heyl, Many-body localization and quantum ergodicity in disordered long-range Ising models, \textit{Phys. Rev. B} \textbf{92}, 134204 (2015).
		
		
		\bibitem{Vn} T. Orell, A. A.Michailidis, M. Serbyn,  and  M. Silveri, Probing the many-body localization phase transition with superconducting circuits, \textit{Phys. Rev. B} \textbf{100}, 134504 (2019).
		
		\bibitem{Vn1} M. Hopjan, and F. Heidrich-Meisner,  Many-body localization from a one-particle perspective in the disordered 1d Bose-Hubbard model, \textit{Phys. Rev. A} \textbf{101}, 063617 (2020).
		
	\bibitem{Supremacy}F. Arute \textit{et al.}, Quantum supremacy using a programmable superconducting processor, \textit{Nature} \textbf{574} 505 (2019). 
		
		
	
		

	
		
	\end{thebibliography}

\begin{thebibliography}{99}
			\bibitem{Zha2020} Zha, C. et al. Ergodic-localized junctions in a periodically-driven spin chain. Preprint at https://arxiv.org/abs/2001.09169.
			\bibitem{Barends2013} Barends, R. et al. Coherent josephson qubit suitable for scalable quantum integrated circuits. \textit{Phys. Rev. Lett.} \textbf{111}, 080502 (2013).
			
			\bibitem{Koch2007} Koch, J., et al. Charge-insensitive qubit design derived from the Cooper pair box. \textit{Phys. Rev. A} \textbf{76}, 042319 (2007).
			
			\bibitem{Yan2019} Yan, Z. et al. Strongly correlated quantum walks with a 12-qubit superconducting processor. \textit{Science} \textbf{364}, 753-756 (2019). 
			
			
			\bibitem{QuTIP} Johansson, J. R., Nation, P. D. \& Nori, F. QuTiP 2: A Python framework for the dynamics of open quantum systems. \textit{Comput. Phys. Commun.} \textbf{183}, 1760 (2012).
			
			
			
			\bibitem{Bayesian} Gelman, A., Carlin, J. B., Stern, H. S., Dunson, D. B., Vehtari, A., \& Rubin, D. B. Bayesian data analysis. (Chapman and Hall/CRC, Boca Raton, 2013).	
		\end{thebibliography}
\end{document}